\documentclass[twocolumn]{emulateapj}
\usepackage[utf8]{inputenc}

\usepackage{natbib}
\usepackage{graphicx}
\usepackage{bibentry}
\usepackage{enumitem}

\def\E{{\cal E}}
\def\F{{\cal F}}

\newcommand{\code}[1]{\texttt{#1}}
\newcommand{\simgt}{\,\hbox{\lower0.6ex\hbox{$\sim$}\llap{\raise0.6ex\hbox{$>$}}}\,}
\newcommand{\simlt}{\,\hbox{\lower0.6ex\hbox{$\sim$}\llap{\raise0.6ex\hbox{$<$}}}\,}

\begin{document}

\title{Sloshing of Galaxy Cluster Core Plasma in the Presence of Self-Interacting Dark Matter}

\author{J.~A. ZuHone}
\affiliation{Harvard-Smithsonian Center for Astrophysics, 60 Garden St., Cambridge, MA 02138, USA}
\author{J. Zavala}
\affiliation{Center for Astrophysics and Cosmology, Science Institute, University of Iceland, Dunhagi 5, 107 Reykjavik, Iceland}
\author{M. Vogelsberger}
\affiliation{Kavli Institute for Astrophysics and Space Research, Massachusetts
Institute of Technology, 77 Massachusetts Avenue, Cambridge, MA 02139, USA}


\begin{abstract}
The ``sloshing'' of the cold gas in the cores of relaxed clusters of galaxies is a widespread phenomenon, evidenced by the presence of spiral-shaped ``cold fronts'' in X-ray observations of these systems. In simulations, these flows of cold gas readily form by interactions of the cluster core with small subclusters, due to a separation of the cold gas from the dark matter (DM), due to their markedly different collisionalities. In this work, we use numerical simulations to investigate the effects of increasing the DM collisionality on sloshing cold fronts in a cool-core cluster. For clusters in isolation, the formation of a flat DM core via self-interactions results in modest adiabatic expansion and cooling of the core gas. In merger simulations, cold fronts form in the same manner as in previous simulations, but the flattened potential in the core region enables the gas to expand to larger radii in the initial stages. Upon infall, the subcluster’s DM mass decreases via collisions, reducing its influence on the core. Thus, the sloshing gas moves slower, inhibiting the growth of fluid instabilities relative to simulations where the DM cross section is zero. This also inhibits turbulent mixing and the increase in entropy that would otherwise result. For values of the cross section $\sigma/m \simgt 1$, subclusters do not survive as self-gravitating structures for more than two core passages. Additionally, separations between the peaks in the X-ray emissivity and thermal Sunyaev-Zeldovich effect signals during sloshing may place constraints on DM self-interactions. 
\end{abstract}

\section{Introduction}

A key ingredient in the standard model of cosmology is the presence of dark matter, which is thought to be composed of one (or several) new particle(s) that have yet not been identified. Observations of stellar motions in galaxies, galaxy motions in clusters, and hot gas temperatures in clusters, among other astrophysical sources of evidence, imply that the total mass of these systems is nearly an order of magnitude more than is visible via electromagnetic radiation. Constraints from Big Bang nucleosynthesis imply that this matter must be in a non-baryonic form. 

In the standard paradigm, dark matter is made of cold and collisionless particles (``cold dark matter'', or CDM), which explains very well the large-scale structure of the universe \citep[][]{Springel2005, Schaye2010, Dubois2014, vogelsberger14a, vogelsberger14b, Schaye2015, Dave2016, McCarthy2018, Springel2018}. However, the model has faced significant difficulties over the last two decades in describing certain aspects of cosmic structure on galactic scales, such as the the presence of low-mass galaxies with cored dark matter profiles, as opposed to the central cusps predicted by CDM \citep[the ``core-cusp problem'';][]{flores94,moore94,moore99}, the wide scatter in halo profile properties \citep[the ``diversity problem'';][]{kuzio2010,oman2015}, and the fact that the most luminous subhalos in the Milky Way are underdense compared with CDM predictions \citep[the ``too big to fail problem'';][]{mbk2011,mbk2012}. Explanations for these discrepancies have been put forward which involve the gravitational effects of the complex baryonic physics that impacts galaxies and their environment: stellar and supernova feedback \citep{mas08,Pontzen2012,Madau2014,Onorbe2015,Read2016,Read2019}, heating from reionisation \citep{efs92,bullock2000,Benson2002,Bovill2009,Sawala2016}, etc. However, it is also possible that these observations are indirect evidence for new dark matter physics. For a recent review on the CDM challenges and range of proposed solutions, see \citet{bullock2017}.

One such class of CDM models which can alleviate these tensions is known as ``self-interacting dark matter'' (hereafter SIDM).\footnote{From here on, ``CDM'' will be used to refer to the standard collisionless DM model.} In these models, the cross section for collisions between DM particles is not negligible but is large enough to produce astrophysically interesting consequences \citep[for a recent review, see][]{tul18}. In these models, DM particles scatter elastically and isotropically with each other, and the collisions conduct heat from the hotter, intermediate regions of the halo to the inner cold cusp. Hence, the density and velocity dispersion profiles of the central halo flatten forming an isothermal core, at least in the early stages before triggering a runaway collapse of the core \citep{koda2011}, the so-called gravothermal collapse in globular clusters \citep{lb1968}. In order to alleviate the aforementioned tensions on galactic scales and avoid the graovthermal collapse, the self-interaction cross-section per unit mass at this mass/velocity scale should be at least $\sigma/m \sim 0.5-1.0$~cm$^2$~g$^{-1}$, at the characteristic velocities of dwarf galaxies \citep{zavala13}. Since the original SIDM idea introduced by \citet{spergel2000} in the context of the CDM challenges, other models have been put forth which include inelastic scattering \citep{todo17a,todo17b,vogelsberger18}, anisotropic scattering \citep{rob17b}, and velocity-dependent SIDM cross sections \citep{vogelsberger12,rob17b}.

SIDM models are constrained more strongly in observations of massive elliptical galaxies and galaxy clusters, where characteristic velocites are higher, in the range 300$-$1000~km~s$^{-1}$. Self-interactions can for instance impact strong lensing signals \citep[$\sigma/m \simlt 0.1$~cm$^2$~g$^{-1}$,][]{meng01}; more recently see also \citet{despali18}, \citealt{rob18b}, and \citealt{andrade2019}; evaporate massive subhaloes below observed abundances \citep[$\sigma/m \simlt 0.3$~cm$^2$~g$^{-1}$,][]{gnedin2001}; 
and reduce the central ellipticity of halos \citep[$\sigma/m \simlt 1$~cm$^2$~g$^{-1}$,][]{peter2013}; 
An examination of combined stellar velocity dispersion and weak lensing measurements in the cores of clusters provided an indirect constraint of $\sigma/m \simlt 0.1$~cm$^2$~g$^{-1}$ \citep{kaplinghat2016}; these observations indicate that the (allowed) radial range that can be affected by self-interactions in galaxy clusters is the inner $\mathcal{O}$(10)~kpc. More recently, \citet{harvey2018} compared oscillations of brightest cluster galaxies (BCGs) in the cores of SIDM halos from cosmological simulations to observed BCG ``wobbles'' to derive a constraint of $\sigma/m \simlt 0.2$~cm$^2$~g$^{-1}$. Such observations imply that the self-interaction cross-section of the DM, if non-zero, must be velocity-dependent, with higher values at smaller halo masses/velocity scales.

Other constraints come from high-speed cluster mergers, where the relative velocity between DM particles can reach several thousand km~s$^{-1}$. The key signatures in this case are the physical separation between baryons and the DM, optical depth arguments, and mass loss from DM interactions. As the self-interaction cross-section increases, separations between DM and stars should increase, as the former experiences drag from collisions and the latter behaves in a collisionless fashion. The most famous example of a high-speed merging cluster is 1E 0657-56, or the ``Bullet Cluster''. Using X-ray and optical observations, a rough limit of $\sigma/m \simlt 1$~cm$^2$~g$^{-1}$ was suggested by \citet{markevitch2004}; by comparing N-body simulations with SIDM to these observations, \citet{randall08} was able to refine this to $\sigma/m \simlt 1.25$~cm$^2$~g$^{-1}$ based on the nonobservation of an offset between the DM mass peak and the galaxy centroid in the western subcluster, and $\sigma/m \simlt 0.7$~cm$^2$~g$^{-1}$ based on mass loss \citep[see also][]{kahlhoefer2014}. Constraints based on other merging clusters obtained using similar methods are within the range of $\sigma/m \sim 1-4$~cm$^2$~g$^{-1}$ \citep[see Table II of][for a summary]{tul18}. Recently, refinements of the method of constraining the self-interaction cross-section in mergers have been investigated in idealized merger simulations (such as those presented in this work) by \citet{rob17a} and \citet{kim17}.

Most of the observational tests of DM collisionality on the cluster scale in terms of spatial separations of DM from baryons have focused on those locked up in stars, which should behave in a collisionless fashion. However, the collisionality of the DM can also be contrasted with that of baryons from the perspective of the X-ray emitting hot plasma, the intracluster medium (hereafter ICM). The collisionality of the ICM in the central cluster region is much higher than that of the DM since the latter is constrained to have a Knudsen number of $\mathcal{O}(1)$. The most famous observational example of this difference is the significant offsets between the gas and DM components in the Bullet Cluster \citep{markevitch2002}, where the cold gas of the core of the western cluster has been pushed out of the DM core by the ram pressure of the surrounding medium, which also strips it and produces a sharp surface brightness discontinuity known as a ``cold front'' \citep[for recent reviews see][]{MV07,ZR16}, where the bright/denser side of the gas is observed to be colder than the fainter/lighter side. Such cold fronts have been observed in other major merging systems. 

There is another class of cluster cold fronts which also appear to depend on the different collisional properties between DM and gas. ``Cool-core'' clusters are relatively relaxed systems which have formed a dense, bright X-ray core, accompanied by a temperature and entropy drop down to the cluster center which have arisen from gas cooling uninterrupted by mergers. In these cores, cold fronts are often observed, laid out in a spiral pattern if multiple fronts are observed. Simulations have shown that these fronts can be produced from interactions of the cool core with small subclusters \citep[e.g.][]{AM06,zuh10,rod12}. As a small cluster or group passes by the core, it gravitationally accelerates both the gas and DM, but these two components separate due to the ram pressure exerted on the gas by the surrounding ICM. Since the gravitational potential is dominated by the DM, after the subcluster passes the cold gas which has been uplifted from the potential minimum falls back toward the center, and begins an oscillatory motion which produces cold fronts. The spiral pattern of these fronts occurs due to the angular momentum transferred to the cold gas from the subcluster if (as is likely) it is not a direct head-on collision. This process and the cold fronts it produces has been dubbed ``sloshing.'' Sloshing cold fronts have been observed in many cool-core clusters \citep[][]{mar00,mvf03,cla04,ghi10,sim10}.

The previous discussion shows that the formation of sloshing cold fronts is crucially dependent on the fact that the gas and DM have different collisionalities \citep{AM06}. The question thus arises as to what effect an increased collisionality of the DM may have on their formation and evolution. Naively, one may suppose that making the DM more collisional would make sloshing less effective, since such collisions would effect a mild form of ``ram pressure'' on the DM, resulting in less of a separation between it and the gas. Thus, the presence of cold fronts in X-ray observations may potentially place a constraint on SIDM. In this work, we seek to investigate the effects of SIDM on the hot plasma of a cool-core cluster undergoing sloshing motions and producing cold fronts using hydrodynamic+DM simulations of a idealized binary cluster merger. We will show in this work that the effects of a non-zero DM cross section on the sloshing process are more complicated than the above simple picture would suggest. 

The structure of this paper is as follows: in Section \ref{sec:methods} we briefly outline the physics employed, the code details, and the setup of the galaxy cluster merger simulations. In Section \ref{sec:results} we present the results of our analysis, and in Section \ref{sec:summary} we summarize these results and present our conclusions. All calculations assume a flat $\Lambda$CDM cosmology with $h$ = 0.71, $\Omega_m$ = 0.27, and $\Omega_\Lambda$ = 0.73 at a redshift of $z$ = 0.

\section{Methods}\label{sec:methods}

\subsection{Basic Physics and Code}

To perform our simulations we use the \code{AREPO} code \citep[][]{springel10} to solve the equations of hydrodynamics and self-gravity. The former employs a finite-volume Godunov method on an unstructured moving-mesh, and the latter is computed via a Tree-PM solver. Our simulations contain three types of Lagrangian mass elements. The gas elements are simulated using the moving-mesh Voronoi tesselation method of \code{AREPO}. The gas is modeled as an ideal fluid with $\gamma = 5/3$ and mean molecular weight $\mu = 0.6$. Our goal in this work is to consider the effects of SIDM on the dynamics and appearance of sloshing cold fronts as seen in X-rays, so we perform our simulations in the simplest possible setting (gravity, hydrodynamics, and DM self-interactions) without the complications of additional physics such as radiative cooling and AGN feedback.

The second set of mass elements are the DM particles, which in the CDM model only interact with each other and with other matter via gravity. DM self-interactions have been incorporated into the \code{AREPO} code after the method of \citet[][]{vogelsberger12}. This implementation within \code{AREPO} has been used in previous works to constrain DM self-interactions at the scale of dwarf galaxies \citep{zavala13,vogelsberger14c}, make predictions for direct DM detection experiments \citep{vogelsberger13}, study their effects on gravitational lensing \citep{diaz2018} and cosmological structure formation \citep{cyrracine16,vogelsberger16,lovell2018}. In this work, we employ this implementation of DM self-interactions, which assumes that the scattering between DM particles is elastic and isotropic. We have chosen a constant value of the DM cross section $\sigma/m$ for all of our simulations. Such models are generally considered to be too simple, and cross sections high enough to explain observations at the galaxy scale are inconsistent with observations at the cluster scale, so velocity-dependent cross section models are preferred \citep[][]{zavala13,kaplinghat2016}. Since our idealized simulations are focused singly on the cluster scale, a single velocity-independent value of $\sigma/m$ for each simulation is sufficient. We investigate the effects of a varying cross section by performing a number of simulations with different values of $\sigma/m$ = 0.0, 0.1, 1.0, 3.0,~and~10.0~cm$^2$~g$^{-1}$. For the $\sigma/m = 0$~cm$^2$~g$^{-1}$ case we simply simulate CDM without the self-interaction model compiled in, but we refer to it by its cross-section value as a shorthand. Given the constraints on the cross section for clusters mentioned above, the value of $\sigma/m = 10$~cm$^2$~g$^{-1}$ is definitively ruled out by observations at the cluster scale, and $\sigma/m = 3$~cm$^2$~g$^{-1}$ is nearly ruled out. These are included here as reference cases.  

The third and final type of mass element we employ are star particles, which will serve as tracers of the stellar material of the cluster. In this work, we only simulate the stars associated with the BCG in the cluster center, which will serve as a useful reference frame for the dynamics of the core region and enable us to examine the relative separations of gas, stars, and DM initiated by the merger. 

\subsection{Initial Conditions}

We use identical initial conditions for our binary cluster merger setup as \citet{AM06} and \citet{zuh10}, which we outline here in brief. 

Our merging clusters consist of a large, ``main'' cluster, and a small infalling subcluster. For the combined density profile (DM+stars) of the main cluster we have chosen a \citet{her90} profile:
\begin{equation}\label{eqn:hernquist}
\rho_{\rm DM+star}(r) = \frac{M_0}{2{\pi}a^3}\frac{1}{(r/a)(1+r/a)^3}
\end{equation}
where $M_0$ and $a$ are the scale mass and length of the combined DM/stellar halo. The Hernquist profile has the same dependence on radius in the center as the well-known NFW profile \citep{NFW97}, $\rho_{\rm DM} \propto r^{-1}$ as $r \rightarrow 0$, but is used here instead because it is more analytically tractable and its mass profile converges as $r \rightarrow \infty$. We also use Equation \ref{eqn:hernquist} for the pure-DM density profile of the subcluster. 

For the stellar component of the BCG, we use an analytical approximation to a deprojected S\'{e}rsic profile given by \citet{mer06}:
\begin{eqnarray}
\rho_{\rm *}(r) &=& \rho_e\exp\left\{-d_{n_s}\left[(r/r_e)^{1/{n_s}}-1\right]\right\}\label{eqn:sersic} \\
d_{n_s} &\approx& 3n_s - 1/3 + 0.0079/n_s,~{\rm for}~n_s \simgt 0.5
\end{eqnarray}
which is a good representation of the stellar mass density profile of elliptical galaxies. We set $n_s = 6$, $\rho_e = 1.3 \times 10^4$~M$_\odot$~kpc$^{-3}$, and $r_e = 175$~kpc, giving a mass $M_{\rm *,BCG} \sim 3 \times 10^{12}$~M$_\odot$, appropriate for a $\sim 10^{15}$~M$_\odot$ cluster \citep[][]{mer06,kravtsov18}. We ignore the stellar mass contribution from other galaxies since we are mainly concerned with the dynamics of the different mass components in the core region. The DM density and mass profiles are then simply the difference of the combined and stellar profiles. 

For the gas density, we use a phenomenological formula which can model cool-core clusters with temperature decreasing towards the cluster center \citep{AM06}:
\begin{equation}
\rho_{\rm gas}(r) = \rho_{g0}\left(1+\frac{r}{a_c}\right)\left(1+\frac{r/a_c}{c}\right)^\alpha\left(1+\frac{r}{a}\right)^\beta,
\end{equation}
with exponents
\begin{equation}
\alpha \equiv -1-n\frac{c-1}{c-a/a_c},~\beta \equiv 1-n\frac{1-a/a_c}{c-a/a_c},
\end{equation}
where $0 < c < 1$ is a free parameter that characterizes the depth of the temperature drop in the cluster center and $a_c$ is the characteristic radius of that drop, or the ``cooling radius''.
We set $n = 5$ in order to have a constant baryon fraction at large radii, and we compute the value of $\rho_0$ from the constraint $M_{\rm gas}/M_{\rm DM} = \Omega_{\rm gas}/\Omega_{\rm DM} = 0.12$. With this density profile and Equation \ref{eqn:hernquist}, the corresponding gas temperature can be derived by imposing hydrostatic equilibrium.

We perform two types of simulations. In one set, we evolve the a cool-core cluster in isolation for a number of different values of the self-interaction cross section to test the effect of DM collisions on the gas properties in the absence of a merger. In the second set, we perform simulations of a merger between the same cool-core cluster and a small subcluster to produce sloshing gas motions and cold fronts. In this set, we also include other simulations where the main cluster is allowed to form a DM core via self-interactions for several Gyr before undergoing the merger, and a simulation where self-interactions are not switched on until shortly before the first core passage of the two clusters. We will describe the rationale for these simulations in more detail in Section \ref{sec:mergers}.

The two clusters are characterized by the mass ratio $R \equiv M_1/M_2 = 5$, where $M_1 = M_0R/(1+R)$ and $M_2 = M_0/(1+R)$ are the masses of the main cluster and the infalling satellite, respectively. The total cluster mass $M_0$ is set to $1.5 \times 10^{15}$~M$_\odot$. To scale the initial profiles for the two subclusters, the combination $M_i/a_i^3$ in Equation \ref{eqn:hernquist} is held constant. For the main cluster, we chose $a_1$ = 600~kpc, $c$ = 0.17, and $a_c$ = 60~kpc, to resemble mass, gas density, and temperature profiles typically observed in real galaxy clusters. In particular, our main cluster closely resembles A2029 \citep[e.g.,][]{vik05}, a hot, relatively relaxed cluster with sloshing in the cool core. The subcluster contains DM only and has a mass density profile given by Equation \ref{eqn:hernquist}, though in Section \ref{sec:appendix} we describe simulations where a BCG stellar component is added to it. The choice of a subcluster without a baryonic component is somewhat unusual, but we use it in this case (as in the previous works) to produce relatively undisturbed cold fronts without significant shocks or turbulence. In future papers we plan to investigate idealized merger simulations with gas-filled halos. 
    
\renewcommand{\arraystretch}{1.5} 
\begin{table}[thp]
\caption{Simulation Parameters\label{tab:sim_params}}
\begin{center}
\begin{tabular}{ll}
\hline
Total DM+Stellar Mass ($M_0$) & \\
\hspace{6mm}Main Cluster & $1.25 \times 10^{15}$~M$_\odot$ \\
\hspace{6mm}Subcluster & $2.5 \times 10^{14}$~M$_\odot$ \\
Scale Radius ($a$) & \\
\hspace{6mm}Main Cluster & 600~kpc \\
\hspace{6mm}Subcluster & 350~kpc \\
BCG S\'{e}rsic Index ($n_s$) & 6 \\
BCG Scale Density ($\rho_e$) & $1.3 \times 10^4$~M$_\odot$~kpc$^{-3}$ \\
BCG Scale Radius ($r_e$) & 175~kpc \\
Cooling Radius ($a_c$) & 60~kpc \\
Temperature Drop Parameter ($c$) & 0.17 \\
Cluster Separation ($d$) & 3.0~Mpc \\
Initial Impact Parameter ($b$) & 0.5~Mpc \\
Initial Relative Velocity ($v_1-v_2$) & 1466~km~s$^{-1}$ \\
\hline
\end{tabular}
\end{center}
\end{table}
\renewcommand{\arraystretch}{1} 

With the characteristics of the clusters thus defined, we may set up the particle and cell properties in the simulations. The DM and star particles all have the same mass. The gas cells are initialized to all have the same mass, though they are allowed to undergo mesh refinement and derefinement during the simulation evolution, so this condition will not remain strictly true in their case as the simulation progresses. For each of the particle/cell positions, a random deviate $u = M(<r)/M_{\rm total}$ is uniformly sampled in the range [0, 1] and the mass profile $M(<r)$ for that particular mass type is inverted to give the radius of the particle/cell from the center of the halo.

\begin{figure*}
\centering
\includegraphics[width=0.8\textwidth]{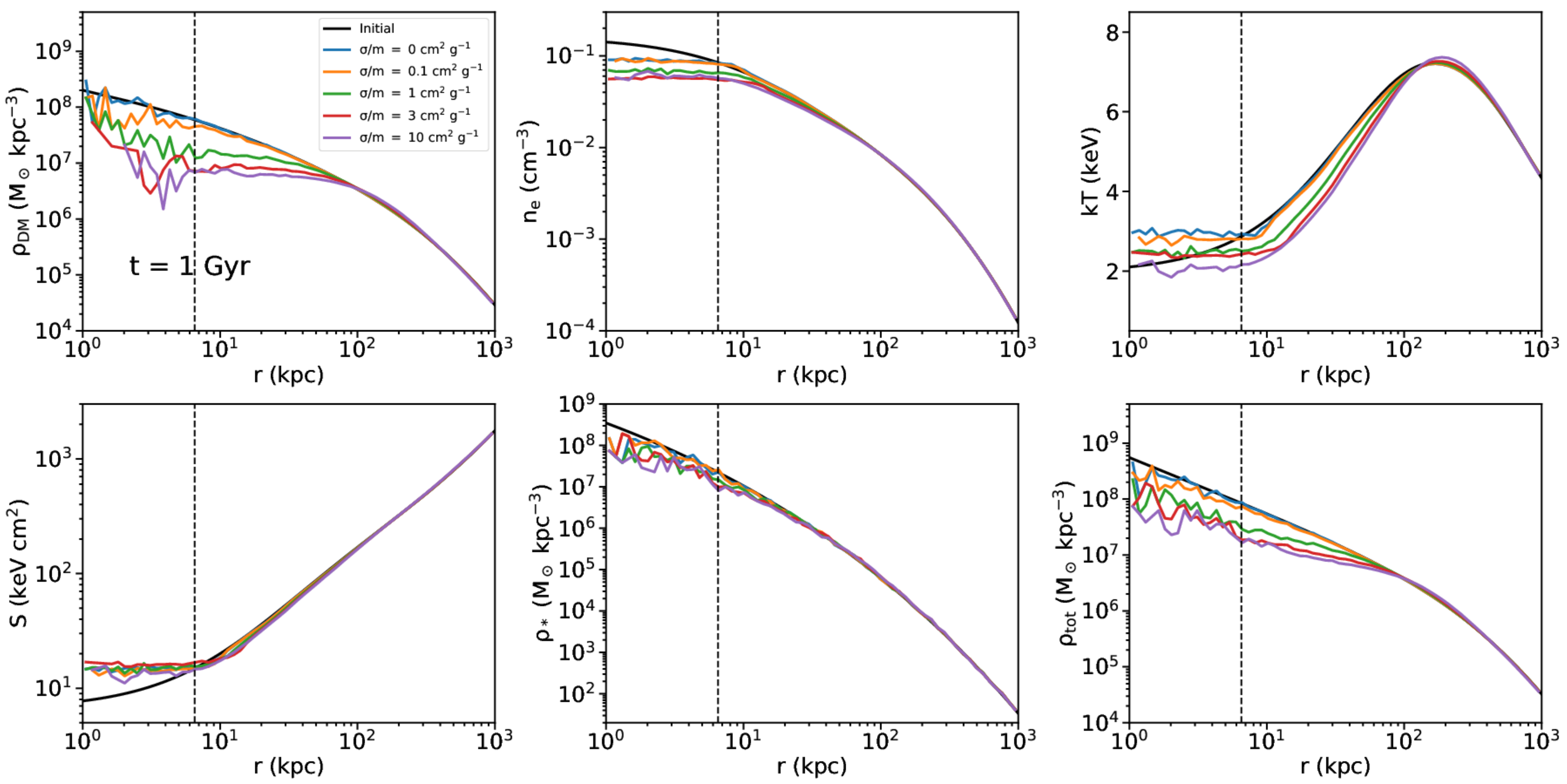}
\includegraphics[width=0.8\textwidth]{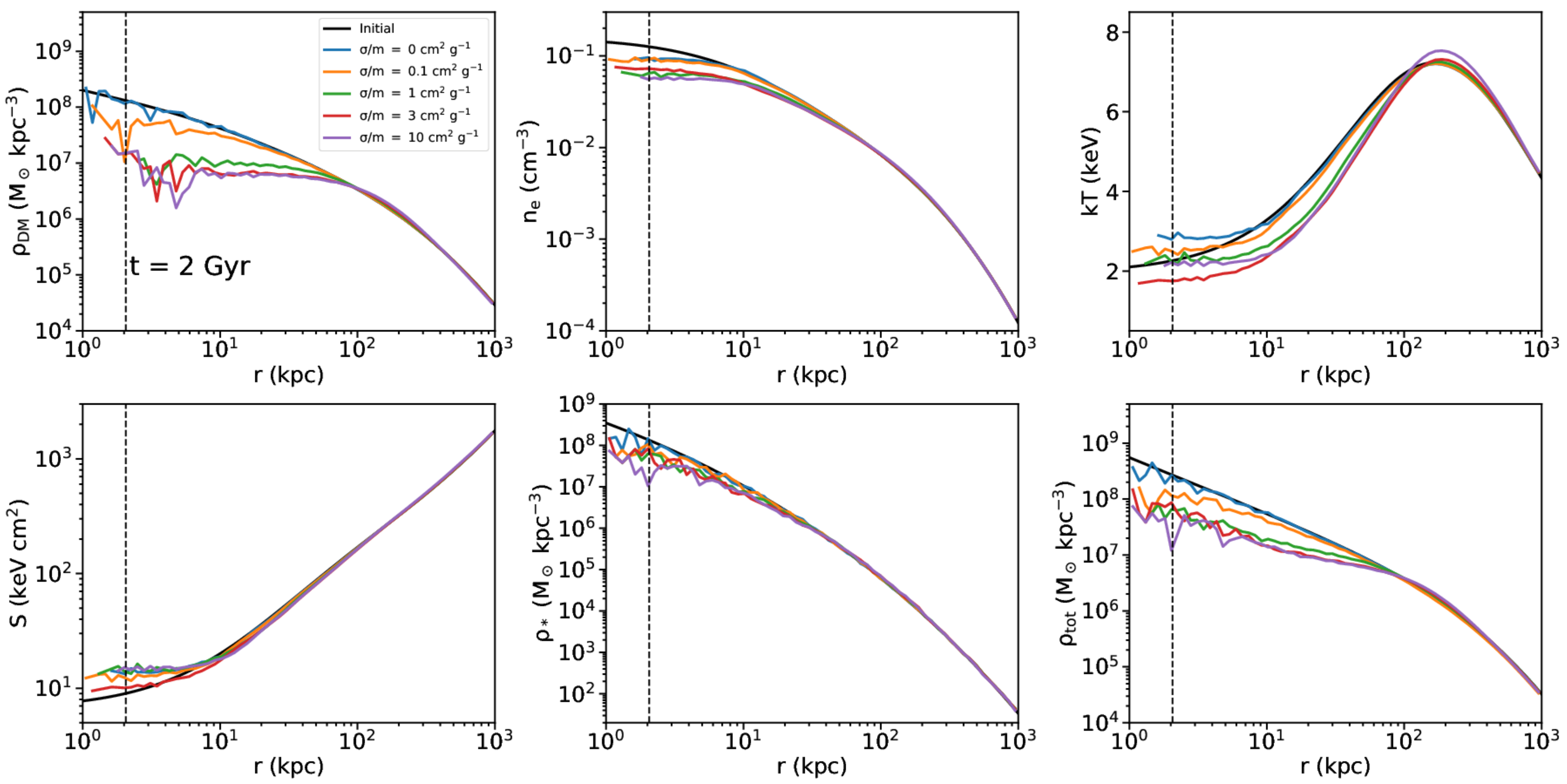}
\includegraphics[width=0.8\textwidth]{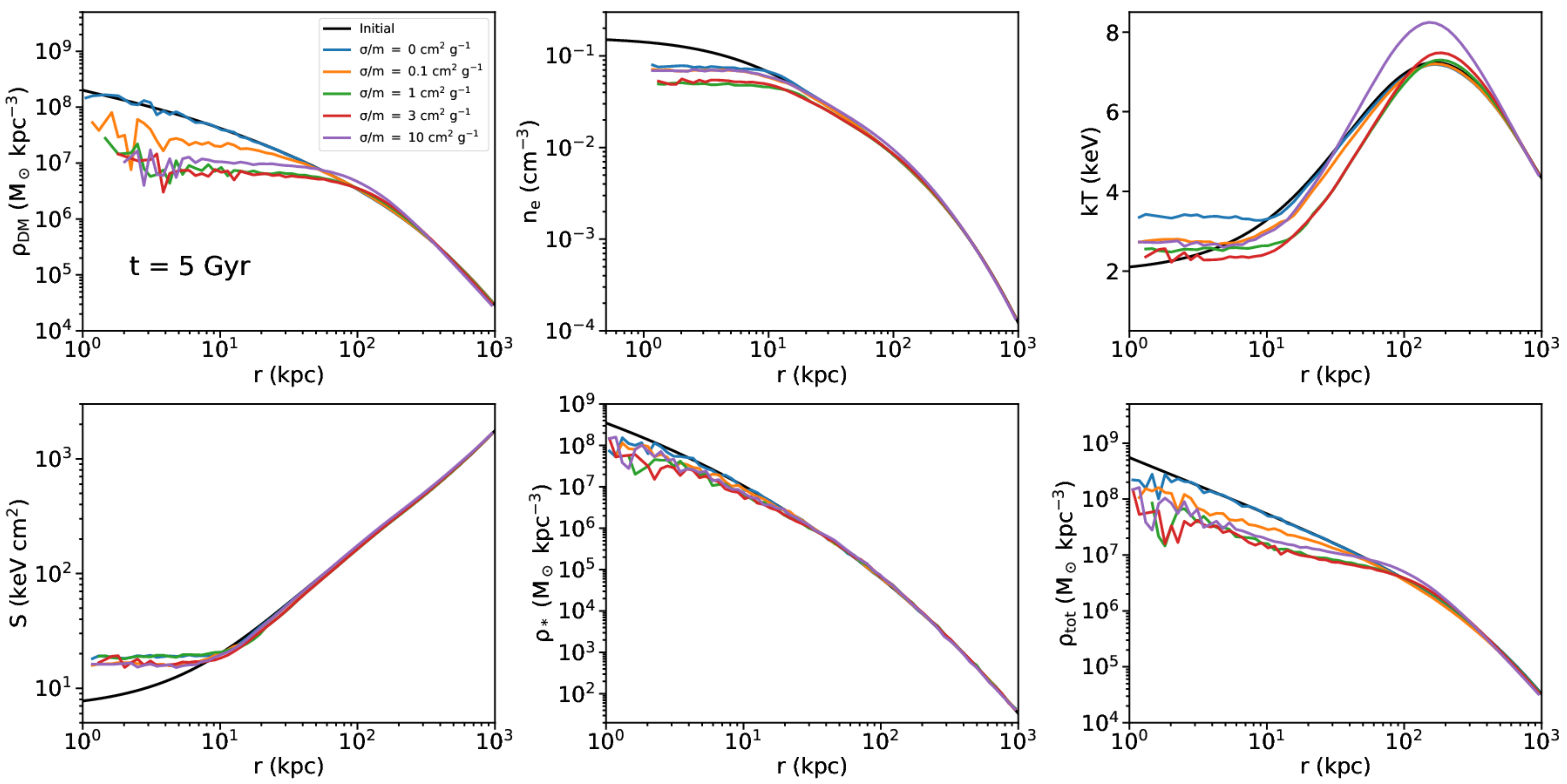}
\caption{Spherically averaged radial profiles of various quantities at $t$ = 1, 2, and 5~Gyr (6 panels at each epoch) for the single-cluster tests for different values of $\sigma/m$ (shown with different colors as given in the legends). For each epoch, the top panels show DM density, electron number density, and gas temperature, and the bottom panels show gas entropy, stellar density, and total (DM+gas+stars) density. The solid dashed lines mark the ``convergence radius'' \citep{power2003} at each epoch.\label{fig:test_single}}
\end{figure*}
    
The gas cells are assigned densities and internal energies from the gas density and temperature profiles, with their initial velocities set to zero in the rest frame of the cluster. For the DM and star particles, their initial velocities are determined using the procedure outlined in \citet{kaz04}, where the energy distribution function is calculated via the Eddington formula \citep{edd16}:
\begin{equation}
\F(\E) = \frac{1}{\sqrt{8}\pi^2}\left[\int^\E_0{d^2\rho \over d\Psi^2}{d\Psi \over \sqrt{\E - \Psi}} + \frac{1}{\sqrt{\E}}\left({d\rho \over d\Psi}\right)_{\Psi=0} \right]
\end{equation}
where $\Psi = -\Phi$ is the relative potential and $\E = \Psi - \frac{1}{2}v^2$ is the relative energy of the particle. We tabulate the function $\F$ in intervals of $\E$ interpolate to solve for the distribution function at a given energy. Given the radius of the particle, particle speeds can then be chosen from this distribution function using the acceptance-rejection method. Once particle radii and speeds are determined, positions and velocities are determined by choosing random unit vectors in $\Re^3$.

The main cluster is evolved in isolation (without self-interactions) for several dynamical times to smooth out initial pressure and density fluctuations of the gas particles. The resulting equilibrium profiles are essentially identical to the initial setup, and these are the initial conditions for the main cluster that we use for all of our simulations.

For the merger simulations, both objects start at a separation of $d$ = 3~Mpc, and with an initial impact parameter $b$ = 500~kpc. The initial cluster velocities are chosen so that the total kinetic energy of the system is set to half of its potential energy, under the approximation that the objects are point masses:
\begin{equation}
E \approx -\frac{1}{2}\frac{GM_1M_2}{d} = -\frac{1}{2}\frac{R}{(1+R)^2}\frac{GM_0^2}{d}
\end{equation} 
So the initial velocities in the reference frame of the center of mass are set to
\begin{equation}
v_1 = \frac{R}{1+R}\sqrt{\frac{GM_0}{d}} ; v_2 = \frac{1}{1+R}\sqrt{\frac{GM_0}{d}}
\end{equation}.
    
All simulations are set within a cubical computational domain of width $L = 40$~Mpc on a side, though for all practical purposes the region of interest is confined to 
the inner $\sim 10$~Mpc.
    
The main cluster has $2.375 \times 10^7$ gas cells, each initially with mass $m_{\rm gas} = 1.14 \times 10^7 M_\odot$, though the mass of these cells is allowed to change slightly during the simulation evolution. The main cluster also has $8.45 \times 10^6$ DM particles with mass $m_{\rm DM} = 1.18 \times 10^8 M_\odot$, and $2.7 \times 10^4$ star particles with mass $m_{\rm star} = 1.18 \times 10^8 M_\odot$. The subcluster contains $2 \times 10^6$ DM particles of the same mass $m_{\rm DM}$. The gravitational softening length for the gas cells and the DM and stellar particles is 2~kpc.

A summary of the simulation parameters can be found in Table \ref{tab:sim_params}.

\section{Results}\label{sec:results}

\subsection{Single Clusters Evolved in Isolation}\label{sec:single}

In the first set of simulations, we evolve the main cluster in isolation for each of the simulated values for the DM cross section $\sigma/m$ in order to determine the effect of the self-interactions on the profiles of the DM, stars, and gas in the absence of a merger. Figure \ref{fig:test_single} shows spherically averaged radial profiles of the DM density, electron number density density, gas temperature, gas entropy (defined as $S = k_BTn_e^{-2/3}$), stellar density, and total density for the three different epochs of $t$ = 1, 2, and 5~Gyr for all of the simulated cross sections in this work. The first two epochs are significant as they bracket the time of first core passage in the subsequent merger simulations. 
    
The black lines in each panel show the initial profile at $t = 0$~Gyr. It should first be noted that for the case with $\sigma/m = 0$~cm$^2$~g$^{-1}$ (CDM, blue curves), the profiles are stable for every epoch with the exception of the inner $\sim$10~kpc, where the gas quantities flatten out due to limited force resolution and Poisson noise. The convergence radius of the halo, as defined by \citet{power2003}, is marked by the vertical dashed line in each plot, and is $\sim$6.5~kpc or less (depending on the epoch), implying that our radial profiles are converged roughly outside this radius. This is also implied by the stability of the profiles for the collisionless simulation. Non-zero values of the DM cross section result in a flattening of the DM core density, which happens more quickly for larger values of $\sigma/m$ but the size of the core at later times tends towards the same for all cross sections \citep[as in][]{rob17a}. At very late times, for the largest cross sections the DM density slowly begins to increase again due to the ``gravothermal catastrophe'' inherent in any self-gravitating system where collisions can carry energy away from the core region of the system i.e. with negative heat capacity \citep{lb1968,kw2000}.

\begin{figure*}
\includegraphics[width=0.98\textwidth]{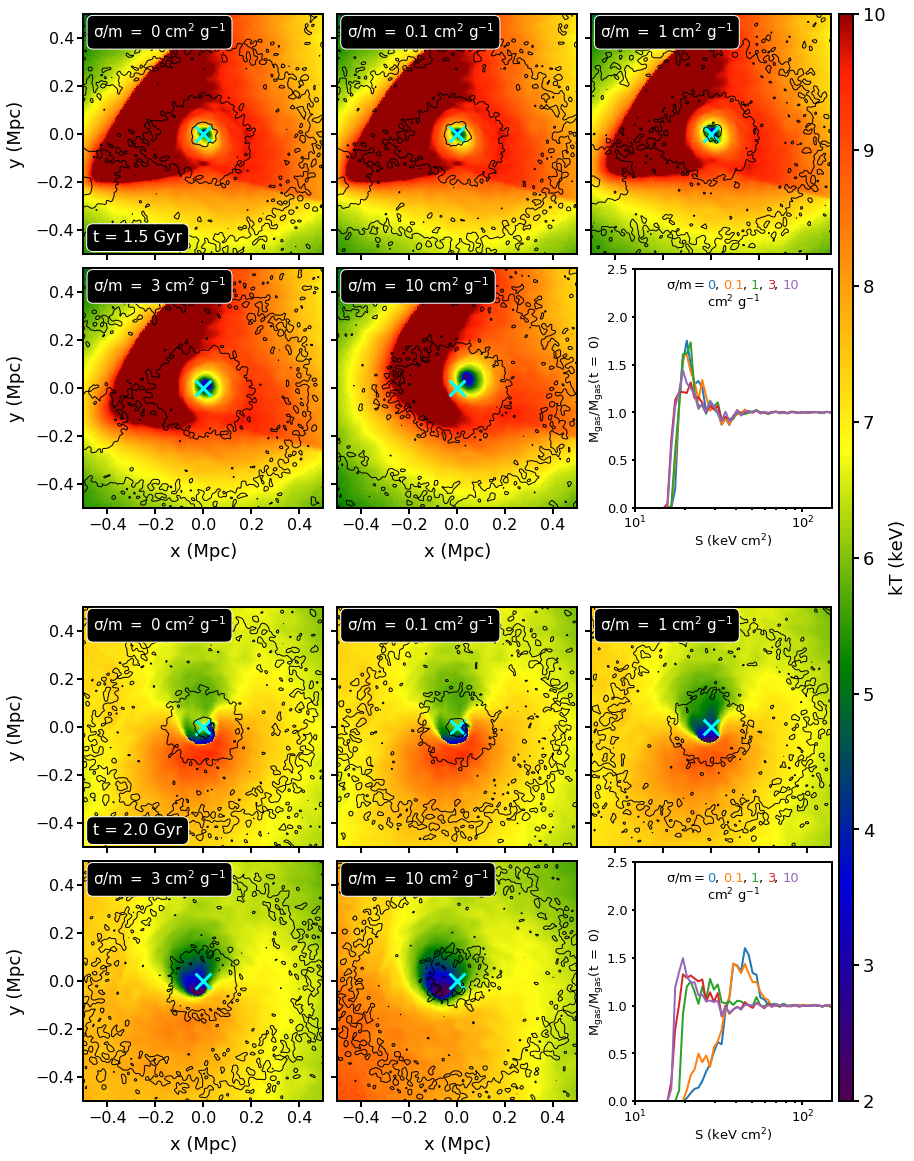}
\caption{Slices through the gas temperature in keV at the epochs $t$ = 1.5 and 2.0~Gyr (top and bottom respectively) for the merger simulations with the 5 different cross sections. Contours are of dark matter density and are spaced logarithmically. The cyan ``$\times$'' marks the position of the center of mass of the BCG. Each panel is 1~Mpc on a side. The bottom-right panel for each epoch shows the ratio of the mass of gas at a given entropy at that epoch to that at the same entropy at $t = 0$ for the different simulations.}
\label{fig:mergers_1.5_2Gyr}
\end{figure*}
    
\begin{figure*}
\includegraphics[width=0.98\textwidth]{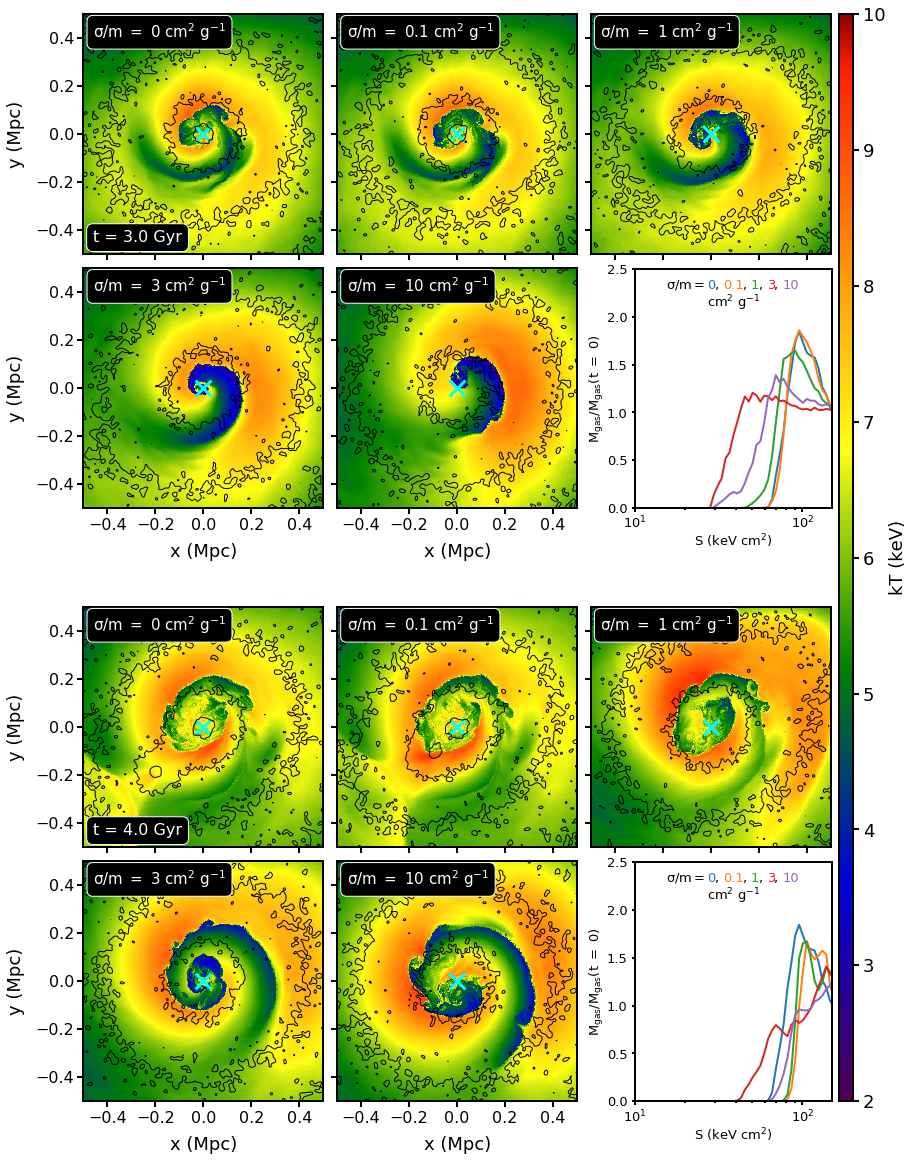}
\caption{Slices through the gas temperature in keV at the epochs $t$ = 3.0 and 4.0~Gyr (top and bottom respectively) for the merger simulations with the 5 different cross sections. Contours are of dark matter density and are spaced logarithmically. The cyan ``$\times$'' marks the position of the center of mass of the BCG. Each panel is 1~Mpc on a side. The bottom-right panel for each epoch shows the ratio of the mass of gas at a given entropy at that epoch to that at the same entropy at $t = 0$ for the different simulations.}
\label{fig:mergers_3_4Gyr}
\end{figure*}
    
Since the flattening of the DM core is a gradual process, the response of the gas to the changing gravitational potential is an adiabatic expansion--the gas density in the core decreases and the temperature decreases (outside of the inner $\sim$10~kpc as noted above). These effects are more pronounced for larger SIDM cross sections. However, the changes in the gas density and temperature are rather small, roughly a factor of $\sim$2 at most, so the system retains its identity as a ``cool-core'' cluster. The gas entropy profile is essentially the same across the simulations, consistent with the assumption that the changes are adiabatic. At larger radii near $r \sim 200$~kpc, where the DM density increases beyond its initial value, an adiabatic {\it compression} of the gas occurs, and the temperature and density in this region increase. This effect is most pronounced for the $\sigma/m = 10$~cm$^2$~g$^{-1}$ simulation. Similar to the gas, the stellar density decreases in the center with increasing $\sigma/m$, slightly flattening as the central potential flattens, though the effect is modest, with the central stellar density decreasing by a factor of $\sim$a few at most. Within the innermost $\sim$30~kpc, the stellar component dominates the total mass density due to the flattening of the DM density profile. 

These results point to an important factor affecting the evolution of our idealized merger simulations which we will describe in the next section--even before the first core passage between the main cluster and the subcluster, DM, gas, and stars are already evolving in ways that may affect our conclusions. Specifically, for a significant DM cross section, the sloshing stage will begin when the main cluster already has a DM core and a slightly colder, more diffuse gas core. It will be important to remember the effects of this evolution in the next section where we examine the merger simulations. 
    
\subsection{Merger Simulations}\label{sec:mergers}

\subsubsection{Visible Appearance of the Cold Fronts and the Sloshing Motions}

We will first describe the visual appearance of the sloshing motions and the cold fronts they produce in the different simulations. Figures \ref{fig:mergers_1.5_2Gyr} and \ref{fig:mergers_3_4Gyr} show slices of gas temperature for four different epochs of the simulations for different values of the DM cross section. Contours of DM density are overlaid on these slices, which are centered on the center of mass of the BCG and focus on the core region. The bottom-right panel of each epoch set in these figures also shows the ratio of the mass of gas at a given entropy at that epoch to that at the same entropy at $t = 0$ for the different simulations. Since all of the simulations start with the same initial condition, this allows us to track how the different DM cross sections affect the evolution of the gas entropy.

At $t = 1.5$~Gyr, shown at the top of Figure \ref{fig:mergers_1.5_2Gyr}, the subcluster has recently made its closest approach to the main cluster center (at $t \approx$~1.35~Gyr). It has compressed and heated gas behind it in a ``sonic wake'' \citep[an effect first noted by][]{AM06} which has a fairly similar appearance in all five simulations. This sonic wake is responsible for transferring angular momentum to the cold gas in the core. For larger values of $\sigma/m$, the subcluster has already become far less centrally concentrated due to collisions, which are enhanced particularly during the core passage. The trajectory of the subcluster is only moderately altered by self-interactions at this point, with the exception of the $\sigma/m$ = 10~cm$^2$~g$^{-1}$ case, where it has been slowed down significantly by the drag force to due to the large number of collisions at core passage. For the largest values of $\sigma/m$ = 3 and 10~cm$^2$~g$^{-1}$, the wake has become detached from the subcluster at this stage. We also note that in the case of $\sigma/m = 10$~cm$^2$~g$^{-1}$ the cool core has already been pushed away from the BCG center by as much as roughly 50~kpc. At this early epoch, the mass distribution of entropy between the simulations is still very similar. 

At $t$ = 2~Gyr (at the bottom in Figure \ref{fig:mergers_1.5_2Gyr}), the process of sloshing has begun in earnest. The evolution of the cold fronts proceeds faster for lower values of the cross section, as evidenced by the presence of a sharper temperature gradient in the images between the cold (blue) and hot (orange) gas in these simulations. In general, for larger cross sections, the temperature of the lowest-entropy gas is colder by about $\sim$1~keV, consistent with the result from Section \ref{sec:single} which showed that the slow transition from DM cusp to core resulted in adiabatic expansion and cooling of the most central gas. The two simulations with $\sigma/m = 0, 0.1$~cm$^2$~g$^{-1}$ have already lost a substantial mass of gas with $S \simlt $30~keV~cm$^2$, presumably due to mixing of hot and cold gas, whereas the other simulations have retained this low-entropy gas. Consistent with this, these two simulations are already showing early signs of the Kelvin-Helmholtz instability (hereafter KHI), as seen in the slice images for those cross sections. 

At $t$ = 3~Gyr (shown at the top of Figure \ref{fig:mergers_3_4Gyr}), the cold fronts are very well-developed in all of the simulations. The spatial extent of the fronts is very nearly the same in all simulations, indicating they travel outward with roughly the same radial velocity. Important differences are present, however. First, the cold fronts in the simulations with larger $\sigma/m$ are still noticeably colder by about $\sim$1~keV. Second, at this epoch, it is more obvious that the simulations with larger cross section are less susceptible to the KHI and correspondingly appear smoother than those with lower cross section. These enhanced KHI in the less collisional simulations result in ``box-shaped'' cold fronts and enhanced turbulence and gas mixing, as noted in previous works \citep[e.g.][]{zuh10,rod12}. We will discuss the reason for this somewhat surprising dependence of the KHI on the DM cross-section in Section \ref{sec:phase_space}. The effect of this reduced mixing is shown in the distribution of entropy at this epoch, as the simulations with $\sigma/m \leq 1$~cm$^2$~g$^{-1}$ have less low-entropy gas than the higher cross sections, as shown in the bottom-right panel of the top part of Figure \ref{fig:mergers_3_4Gyr}). However, the trend is somewhat reversed in the extreme case of $\sigma/m = 10$~cm$^2$~g$^{-1}$, since it has lost more low-entropy gas than the $\sigma/m = 3$~cm$^2$~g$^{-1}$ case, and also appears slightly more susceptible to KHI. It should be noted that in this simulation the flattening of the potential is most extreme, and the modest stabilizing effect against KHI provided by the gravitational force is greatly reduced here.

At later times, $t \simgt 4$~Gyr (shown at the bottom of Figure \ref{fig:mergers_3_4Gyr}), the outermost cold fronts have traveled out to a radius where the density profile of the DM is essentially identical across the simulations with varying $\sigma/m$, and thus the subsequent evolution is similar in appearance. The colder gas in the simulations with larger cross section has persisted even to this later time. At this epoch, in all simulations KHI rolls appear at the cold front surfaces. It is also around this epoch, in the simulations with $\sigma/m \leq 1$~cm$^2$~g$^{-1}$, that the subcluster makes a second core passage, moderately heating the core once more and driving a shock front. In the $\sigma/m$ = 3 and 10~cm$^2$~g$^{-1}$ simulations, the subcluster DM has been completely evaporated into the surrounding main cluster DM via collisions by this time, which will be discussed in more detail in Section \ref{sec:dm_evap}. By this time, all simulations have lost significant amounts of low-entropy gas due to KHI and turbulent mixing.
    
\begin{figure}
\centering
\includegraphics[width=0.47\textwidth]{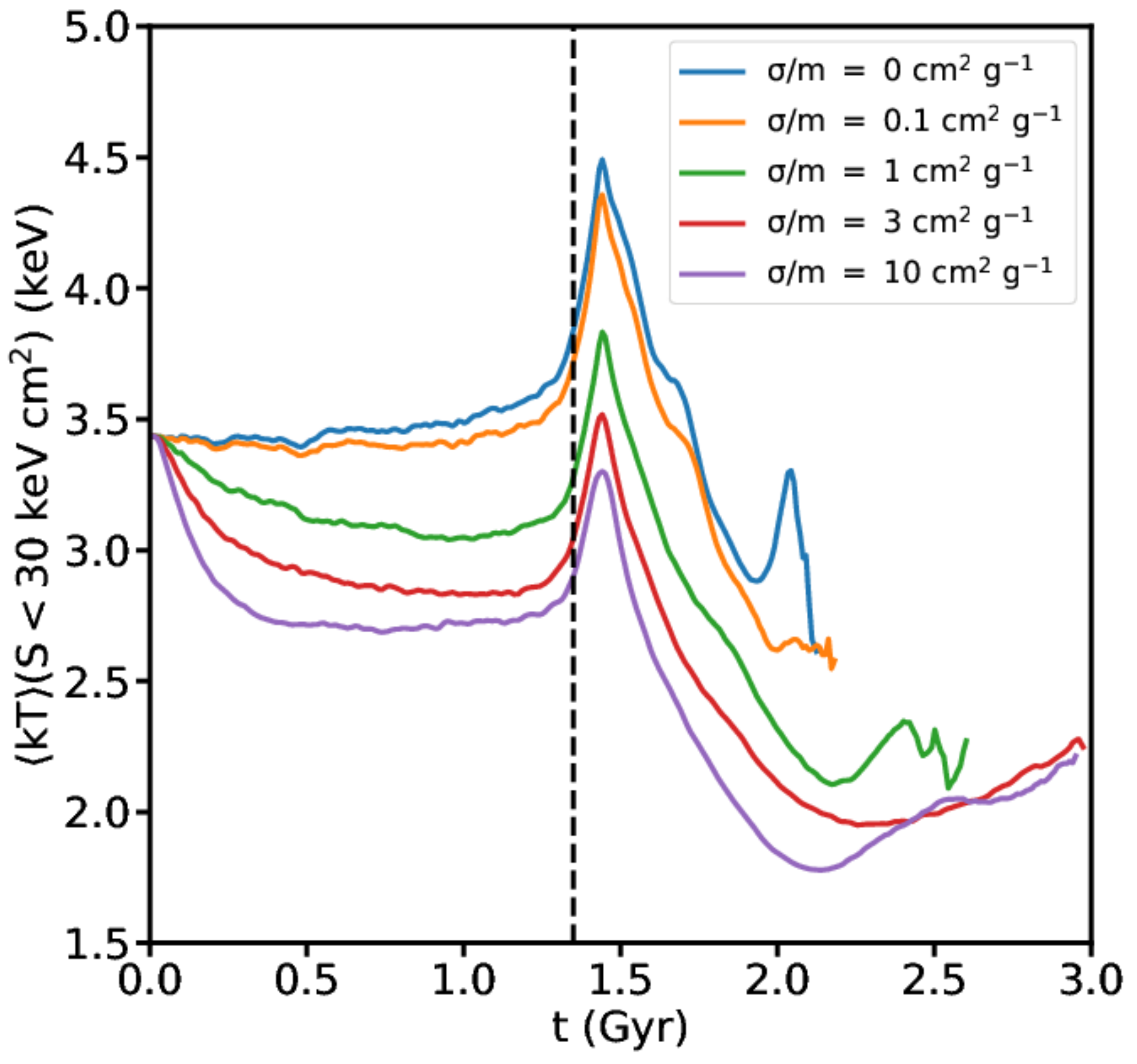}
\caption{Evolution of the mass-weighted average temperature of gas with $S \leq 30$~keV~cm$^2$ for the simulations with different $\sigma/m$. The vertical dashed line marks the approximate epoch of core passage at $t = 1.35$~Gyr.}
\label{fig:temp_evol_sig}
\end{figure}
    
\begin{figure*}
\centering
\includegraphics[width=0.95\textwidth]{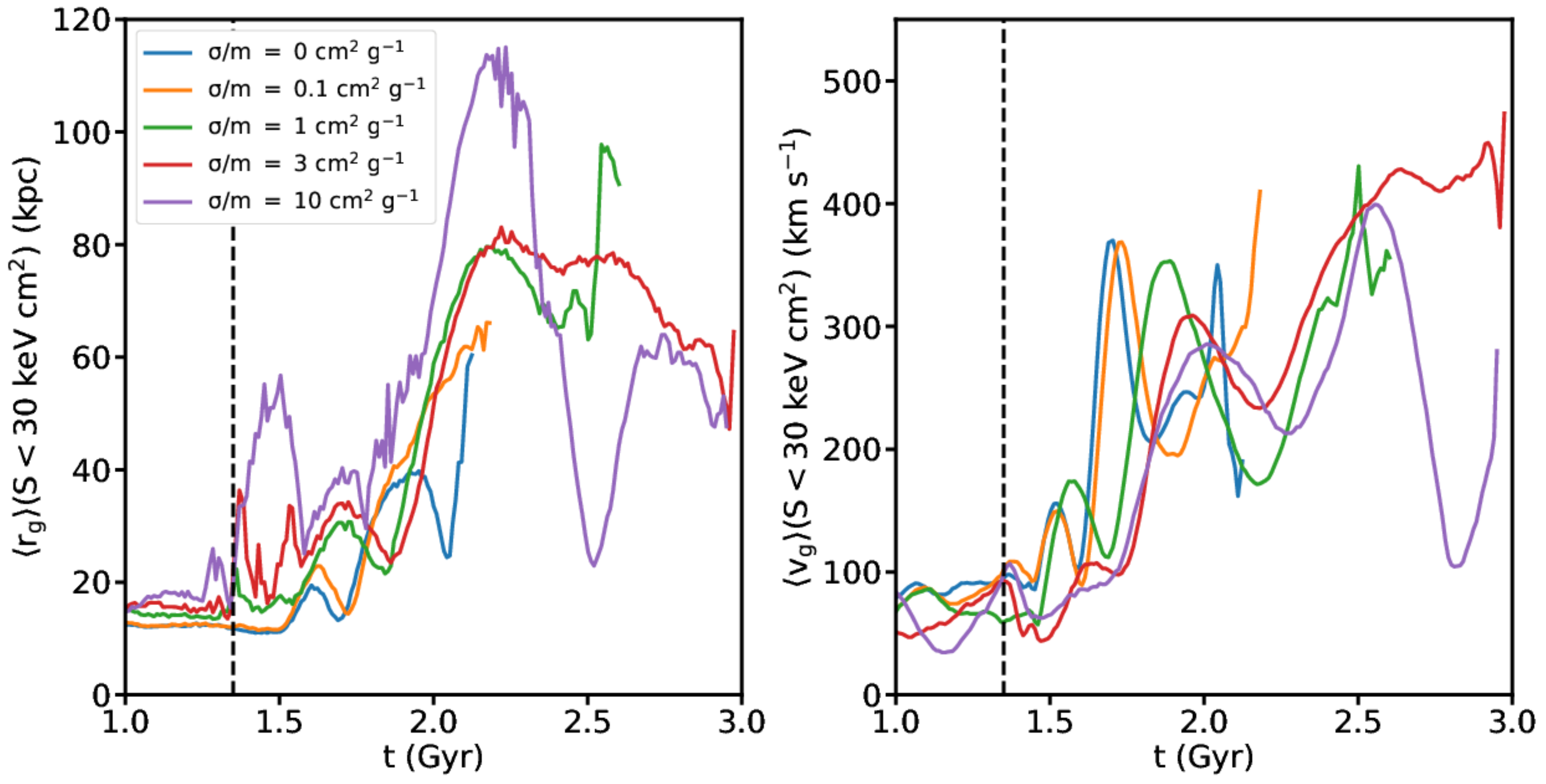}
\caption{Evolution of the mass-weighted average radius and velocity (with respect to the center of mass frame of the BCG) of gas with $S \leq 30$~keV~cm$^2$ for the simulations with different $\sigma/m$. The vertical dashed line marks the approximate epoch of core passage at $t = 1.35$~Gyr.}
\label{fig:phase_evol_sig}
\end{figure*}
    
The effect of the temperature decrease of the core before the core passage due to the adiabatic expansion driven by DM collisions, which persists in the subsequent evolution of the cold fronts, can be seen more clearly in Figure \ref{fig:temp_evol_sig}. In each simulation with self-interactions, the temperature decreases and levels out to a new value before the core passage. This happens more quickly for larger values of $\sigma/m$, and the minimum temperature is also lower. The overall decrease in temperature in the core in the SIDM simulations remains throughout the subsequent evolution, although all simulations show a similar evolution pattern in Figure \ref{fig:temp_evol_sig}. This lower temperature in the SIDM case is somewhat artificial and is a consequence of our idealized setup--in Section \ref{sec:alt_scen} we will investigate how much our conclusions depend on it.

\subsubsection{Phase Space Trajectories of the Cold Gas}\label{sec:phase_space}

From the results above, the behavior of the coldest gas in the core is clearly strongly dependent on the effect that self-interactions have on the gravitational potential in the core. Figure \ref{fig:phase_evol_sig} shows the evolution of the average radius and velocity of the lowest-entropy gas in the cluster core (the averages are taken over all gas with $S \leq 30$~keV~cm$^2$) with respect to the center of mass frame of the BCG. Both quantities increase with time, though superimposed on this increase is an oscillatory motion as the gas sloshes back and forth in the potential. It should be noted that it is {\it not} the case that the same low-entropy gas is gradually rising with radius, which would violate the Schwarzschild stability condition $dS/dr > 0$. Instead, the average entropy of this gas within $S \leq$~30~keV~cm$^2$ is increasing within this limit as cold, low-entropy gas mixes with hot, high-entropy gas from larger radii and the entropy of the core as a whole gradually rises. 

The time period from the core passage at $t \approx 1.35$~Gyr up to $t \sim 2$~Gyr is crucial for the development of the cold fronts. In general, for larger values of the DM cross section, the coldest gas is able to rise to larger radii during this period. This difference in radial extent is significant-- by $t \sim 2$~Gyr, the lowest-entropy gas has risen to only $\sim$40~kpc in the $\sigma/m$ = 0~cm$^2$~g$^{-1}$ simulation, but in the $\sigma/m$ = 1 and 3~cm$^2$~g$^{-1}$ simulations it has risen to $\sim$60~kpc, and in the $\sigma/m$ = 10~cm$^2$~g$^{-1}$ simulation it has risen to $\sim$70~kpc. Increasing the DM cross section has the opposite effect on the speed of the cold gas with respect to the BCG rest frame. This speed tends to be {\it slower} during this period as $\sigma/m$ increases. This effect is similarly dramatic--at $t \sim 1.75$~Gyr the speed of the cold gas in the $\sigma/m$ = 0 and 0.1~cm$^2$~g$^{-1}$ simulations is $\sim$400~km~s$^{-1}$, but in the higher cross section simulations the speed is $\sim$100~km~s$^{-1}$.
    
The explanation for these apparently contradictory behaviors is subtle but straightforward. The flattening of the DM density in the core region which occurs due to DM self-interactions leads to a flattening of the gravitational potential. It is therefore easier for the ram pressure of the surrounding medium to push the gas core out of the DM core towards larger radii against the decreased gravitational force. 

The decrease in core gas speed with increasing cross section is due to the fact that as the subcluster falls into the main cluster, its mass is further reduced by frequent high-velocity collisions with particles from the main cluster's DM during its infall. These interactions are most significant during the short interval of time near the core passage (where the ambient density, and thus the scattering rate, is higher). Thus, the core gas of the main cluster experiences a reduced gravitational acceleration from the passing subcluster, and the sonic wake which is formed by the subcluster and transfers angular momentum to the cold gas is weakened. The fact that this gas is pushed to a larger radius has little effect on its velocity, since the gravitational potential gradient in this region is considerably reduced due to the flattening of the core. This effect is illustrated in more detail in Section \ref{sec:alt_scen}.

Though this gas is moving slower, it nevertheless reaches larger radii than in the simulations with a lower DM cross section, since in the latter simulations in the same time frame it has already reached its peak radius and fallen back into the center. The slower increase of velocity of the cold gas for large cross sections explains why in these simulations KHI appears to be suppressed until later times--the decreased velocity shear across the cold front surface results in an increased growth time for the development of KHI. In general, the slower motions and inhibited KHI associated with the larger cross sections results in less turbulent mixing of the cold gas with hotter gas, as is seen in Figure \ref{fig:phase_evol_sig} by the fact that the lower-entropy gas persists longer in these simulations. 

\begin{figure}
\centering
\includegraphics[width=0.47\textwidth]{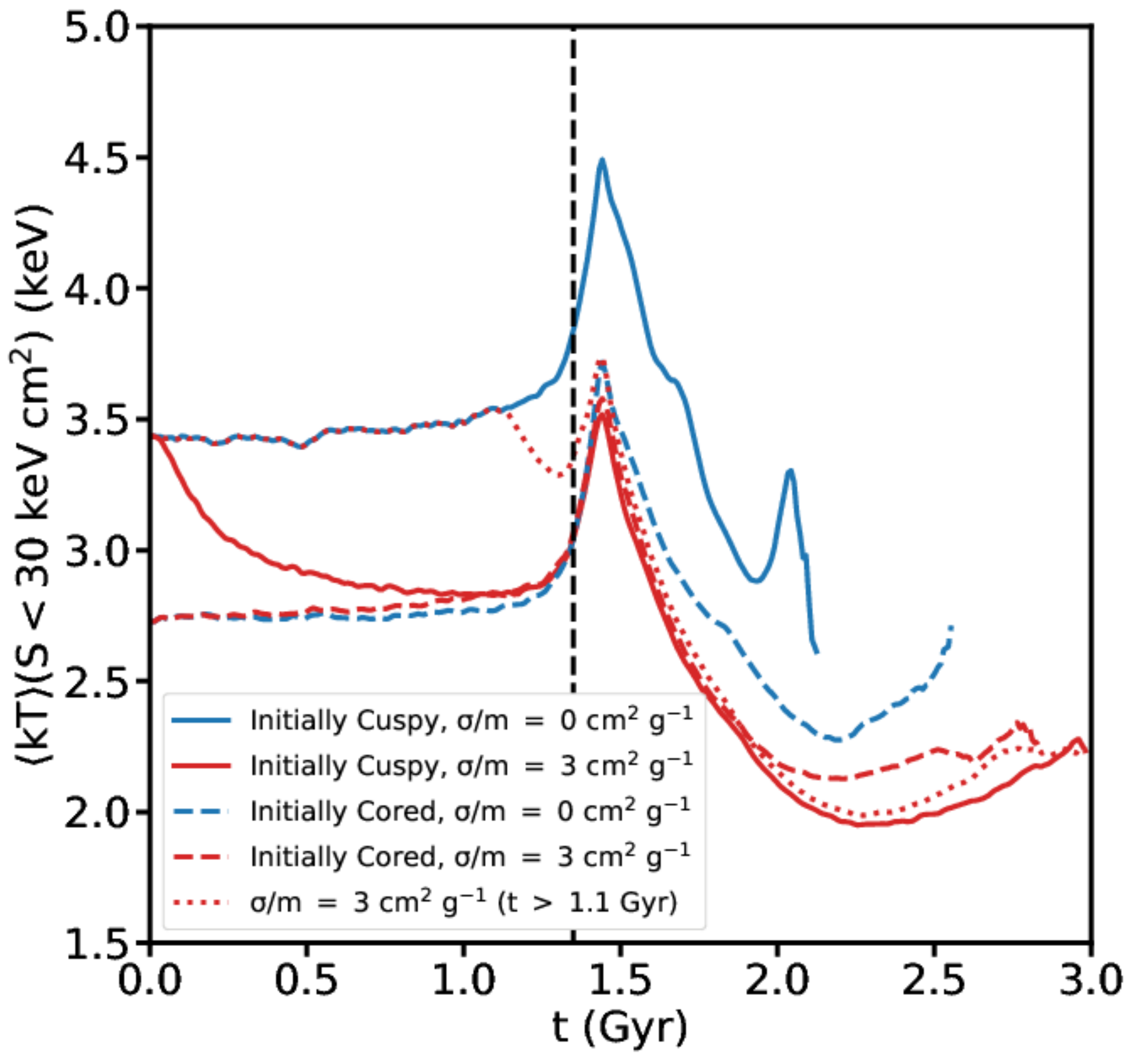}
\caption{Evolution of the mass-weighted average temperature of gas with $S \leq 30$~keV~cm$^2$ for the simulations which test alternative scenarios. The vertical dashed line marks the approximate epoch of core passage at $t = 1.35$~Gyr.}
\label{fig:temp_evol_alt}
\end{figure}

\begin{figure*}
\includegraphics[width=0.98\textwidth]{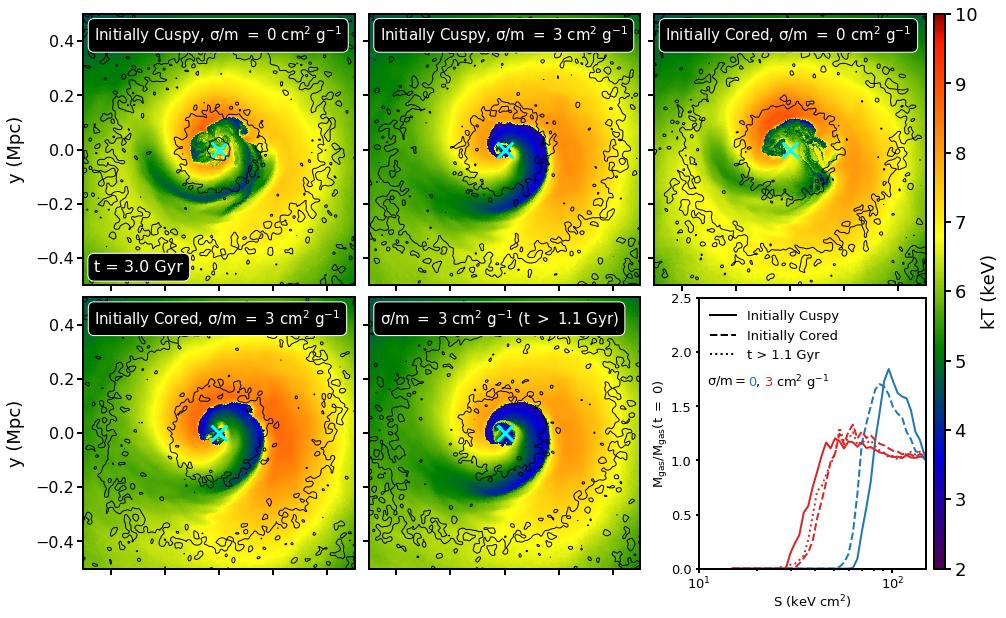}
\caption{Slices through the gas temperature in keV at the epoch $t$ = 3.0~Gyr for the merger simulations in which we test alternative scenarios. Contours are of dark matter density and are spaced logarithmically. The cyan ``$\times$'' marks the position of the center of mass of the BCG. Each panel is 1~Mpc on a side. The bottom-right panel for each epoch shows the ratio of the mass of gas at a given entropy at that epoch to that at the same entropy at $t = 0$ for the different simulations.}
\label{fig:mergers_alt_3Gyr}
\end{figure*}
        
We note that the minor merger we have simulated is not sufficient to result in measurable offsets between the DM core and the BCG--any observable separation would require a direct hit of the subcluster instead of a large impact parameter and possibly a smaller mass ratio (more equal masses) between the two components \citep[see][for a detailed analysis of the required conditions for such separations]{kim17}. In a future paper we will examine such separations between gas, DM, and stars in major mergers.  

\subsubsection{Testing Alternative Scenarios}\label{sec:alt_scen}

In the last two sections, it was determined that the main effects of SIDM on sloshing motions and cold fronts in a cluster core is the flattening of the potential well of the main cluster and the stripping of mass of the subcluster during infall. Since the cluster DM cores are already softening due to self-interactions from the very beginning of the simulation, the conclusions we draw from our merger simulations may depend in a crucial way on this evolution. In this section, we will describe the results of several other simulations we have run to test the robustness of our conclusions to variations in the pre-merger evolution.

The simplest alternative simulation is one where the main cluster halo is not initially cusp-shaped in the core region but already has a density profile that flattens out toward the center. For this simulation we have taken the single-cluster simulation with $\sigma/m = 3$~cm$^2$~g$^{-1}$ and taken its state at $t = 5$~Gyr to be the state of the main cluster at the beginning of the merger simulation. Importantly, we also run the subcluster in isolation for several Gyr so that it too develops a flat DM core. We have run two versions of this simulation, one without any self-interactions for the entire simulation and another where $\sigma/m = 3$~cm$^2$~g$^{-1}$. In theory, this simulation should be very similar to the $\sigma/m = 3$~cm$^2$~g$^{-1}$ case, since in that simulation the cored halo has already established itself before core passage. Figure \ref{fig:temp_evol_alt} shows the evolution of the temperature of the lowest-entropy gas for these simulations (dashed lines) compared to the default versions which begin with cuspy profiles (solid lines). Just before core passage, the temperatures of the lowest-entropy gas are identical in the cored simulations to the original $\sigma/m = 3$~cm$^2$~g$^{-1}$ simulation, and the subsequent evolution of this temperature is very similar. 

However, the subsequent appearance of the cold fronts themselves nevertheless still depends on whether or not the DM is undergoing self-interactions, regardless of the shape of the inner DM density profile. Figure \ref{fig:mergers_alt_3Gyr} shows the appearance of the cold fronts at $t = 3$~Gyr for the DM cusp and DM core simulations with the different cross sections. In both simulations without self-interactions, the cold fronts appear very similar--both have been disturbed by KHI. In both simulations with $\sigma/m = 3$~cm$^2$~g$^{-1}$, the cold fronts appear much smoother. Though the temperature of the lowest-entropy gas is similar between the two simulations with initially flat DM profiles (from Figure \ref{fig:temp_evol_alt}), the overall temperature of the core is hotter in the simulation with $\sigma/m = 0$~cm$^2$~g$^{-1}$ due to the enhanced KHI driving small-scale turbulence and mixing of hot and cold gas phases. The lower-right panel in Figure \ref{fig:mergers_alt_3Gyr} confirms this by showing that the loss of low-entropy gas due to turbulent mixing is driven essentially exclusively by the presence of self-interactions and not the shape of the core potential. 

We also performed another simulation where the DM self-interactions (with $\sigma/m = 3$~cm$^2$~g$^{-1}$) were not switched on until $t$ = 1.1~Gyr, which is right before the core passage at $t \approx 1.35$~Gyr. Though this is a very artificial setup, it avoids the evolution of the gas and DM properties that occur due to self-interactions within the main cluster alone during the period of the subcluster's initial approach. The dotted red line in Figure \ref{fig:temp_evol_alt} shows the temperature of the lowest-entropy gas in this simulation, which begins to adiabatically cool right after the self-interactions are switched on at $t$ = 1.1~Gyr. Its subsequent evolution is nearly identical to the other two simulations with self-interactions, and the appearance of the cold fronts at later times is also very similar to that in these simulations, as seen in Figure \ref{fig:mergers_alt_3Gyr}.

\begin{figure*}
\centering
\includegraphics[width=0.98\textwidth]{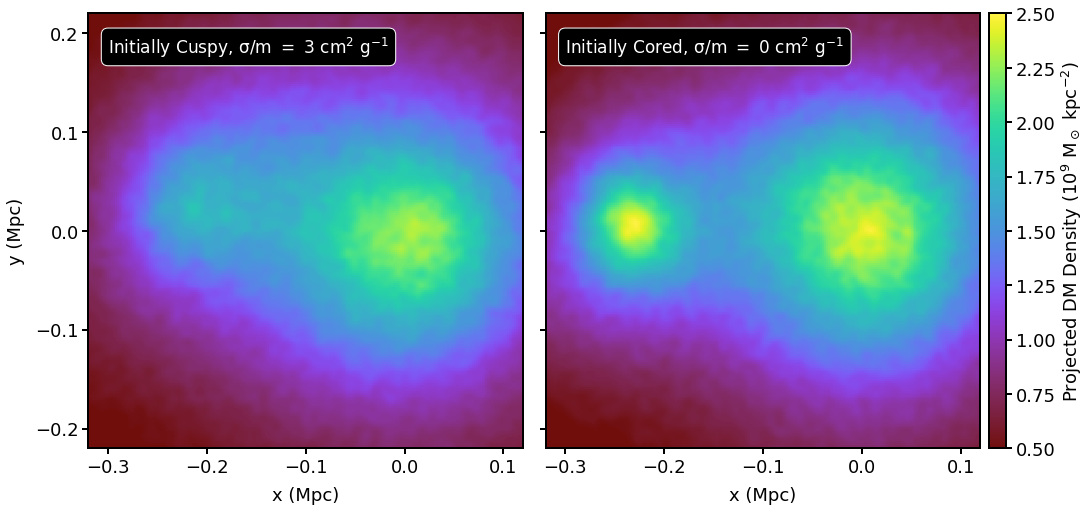}
\caption{Projected DM density at the epoch $t$ = 1.4~Gyr, immediately after core passage, for the initially cuspy simulation with $\sigma/m = 3$~cm$^2$~g$^{-1}$ and the initially cored simulation with $\sigma/m = 0$~cm$^2$~g$^{-1}$. Each panel is 400~kpc on a side.}\label{fig:dm_proj_compare}
\end{figure*}

These results point to the fact that during a merger, self-interactions are a critical effect beyond simply creating DM cores. This is illustrated clearly in Figure \ref{fig:dm_proj_compare}, which shows the projected DM density in the core region shortly after core passage for the initially cuspy simulation with $\sigma/m = 3$~cm$^2$~g$^{-1}$ (left panel), and the initially cored simulation with $\sigma/m = 0$~cm$^2$~g$^{-1}$ (right panel). We find that in the ``cuspy SIDM'' simulation the enclosed mass within a $\sim$50(100)~kpc radius has been reduced by nearly $\sim$40(20), while in the ``cored CDM'' simulation the subcluster has essentially the same enclosed mass at these radii. The decrease of mass of the subcluster will both weaken the sonic wake which transfers angular momentum to the cold gas and decrease the acceleration on the main cluster core itself.
    
\begin{figure*}
\centering
\includegraphics[width=0.95\textwidth]{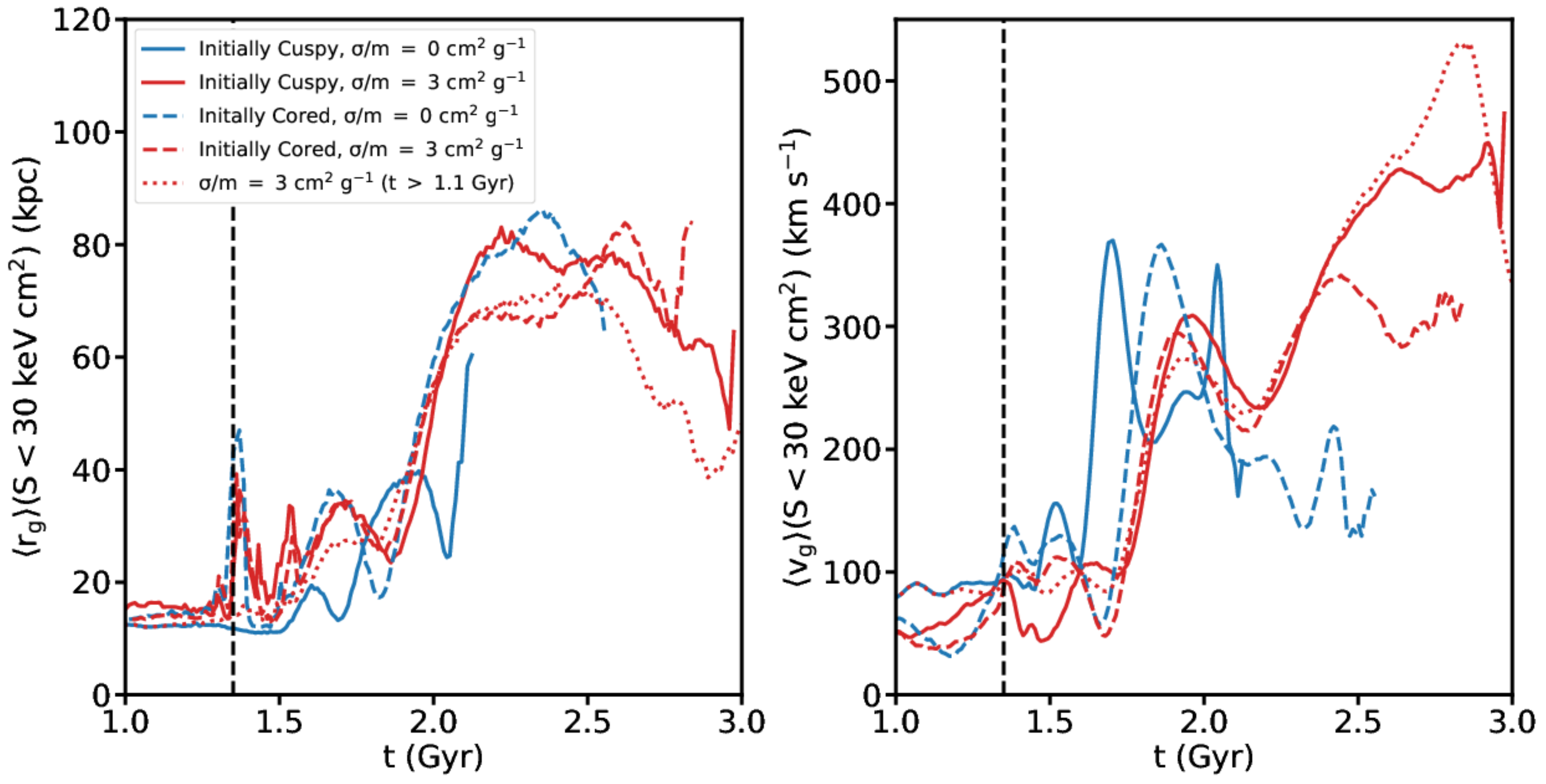}
\caption{Evolution of the mass-weighted average radius and velocity (with respect to the center of mass frame of the BCG) of gas with $S \leq 30$~keV~cm$^2$ for the simulations which test alternative scenarios. The vertical dashed line marks the approximate epoch of core passage at $t = 1.35$~Gyr.}
\label{fig:phase_evol_alt}
\end{figure*}

This is illustrated by Figure \ref{fig:phase_evol_alt}. In all of the simulations in which the main cluster either begins with or develops a flat DM core, the potential is very shallow and easy for the cold, low-entropy gas to climb to larger radii (left panel of Figure \ref{fig:phase_evol_alt}), as was previously noted in Section \ref{sec:phase_space}. The cold gas climbs out to nearly the same radius by $t \sim 2.1-2.2$~Gyr. The behavior of the velocity of this cold gas is somewhat different, however. Though the increase in velocity in the simulation with $\sigma/m = 0$~cm$^2$~g$^{-1}$ and a flat DM core is delayed with respect to the same simulation with a cuspy DM core, the increase in velocity is even slower for the simulations with $\sigma/m = 3$~cm$^2$~g$^{-1}$, which all exhibit similar behavior regardless of the DM core shape (right panel of Figure \ref{fig:phase_evol_alt}). These slower velocities result in longer growth times for KHI and noticeably smoother cold fronts. Thus, the crucial factor in the inhibition of KHI and the resulting smoothness of cold fronts and longevity of low-entropy gas in these simulations is the decrease in the mass of the subcluster due to DM self-interactions.
    
These results assumed that the DM halos of the main cluster and subcluster are in the early evolutionary stages where the cores are softening due to self-interactions. If a merger occurs during a later stage of evolution where DM cusps are developing due to gravothermal catastrophe (trigerred by DM self-interactions), the results may be somewhat different. At first, it may appear that this situation would be similar to the case of collisionless DM, since both the main cluster and subcluster will have deeper gravitational potential wells. However, it should still be expected that the subcluster will experience mass loss due to high-speed collisions of its own DM particles with those of the main cluster. This will result in a weaker influence of the subcluster on the main cluster core, and since the latter's gravitational potential well will be steeper, this will result in reduced sloshing motions.

\subsubsection{Effect of the Core Passage on the Main Cluster DM Core}\label{sec:core_passage}
    
In our simulations, the main cluster's DM core undergoes a transformation from a higher-density cuspy profile 
to a lower-density cored profile due to DM self-interactions, 
whether during a merger or in isolation. During the first core passage, the DM particles from the main cluster core come into contact with those from the subcluster at high relative speed. The high density of the subcluster and the high relative speed of the collisions both increase the number of collisions and the likelihood that these collisions can eject DM particles from the core, further reducing the core density. Figure \ref{fig:before_after_density} shows this effect. The left panel shows the spherically averaged DM density profile for the single-cluster and merger simulations for three different values of the DM cross section, at $t = 1$~Gyr, shortly before the first core passage of the subcluster. The density profiles are the same between the single-cluster and merger simulations at this epoch, as expected. The right panel shows the same profiles after the core passage, at $t = 2$~Gyr. When there are no collisions ($\sigma/m = 0$~cm$^2$~g$^{-1}$), the density profile is unchanged. For $\sigma/m$ = 1 and 3~cm$^2$~g$^{-1}$, the increased number of collisions has in fact decreased the density of the core, but only slightly, by roughly a factor of $\sim$1.5-2 at most. It should be noted that this minor change is consistent with an encounter with a subcluster 5 times less massive at a relatively high initial impact parameter of 0.5~Mpc. Mergers with more equal masses and smaller impact parameters will result in stronger effects on the DM core after the first core passage--exploration of these scenarios will be the subject of a follow-up paper.
        
\subsubsection{Disappearance of the Subcluster Due to DM Self-Interactions}\label{sec:dm_evap}

\begin{figure*}
\centering
\includegraphics[width=0.98\textwidth]{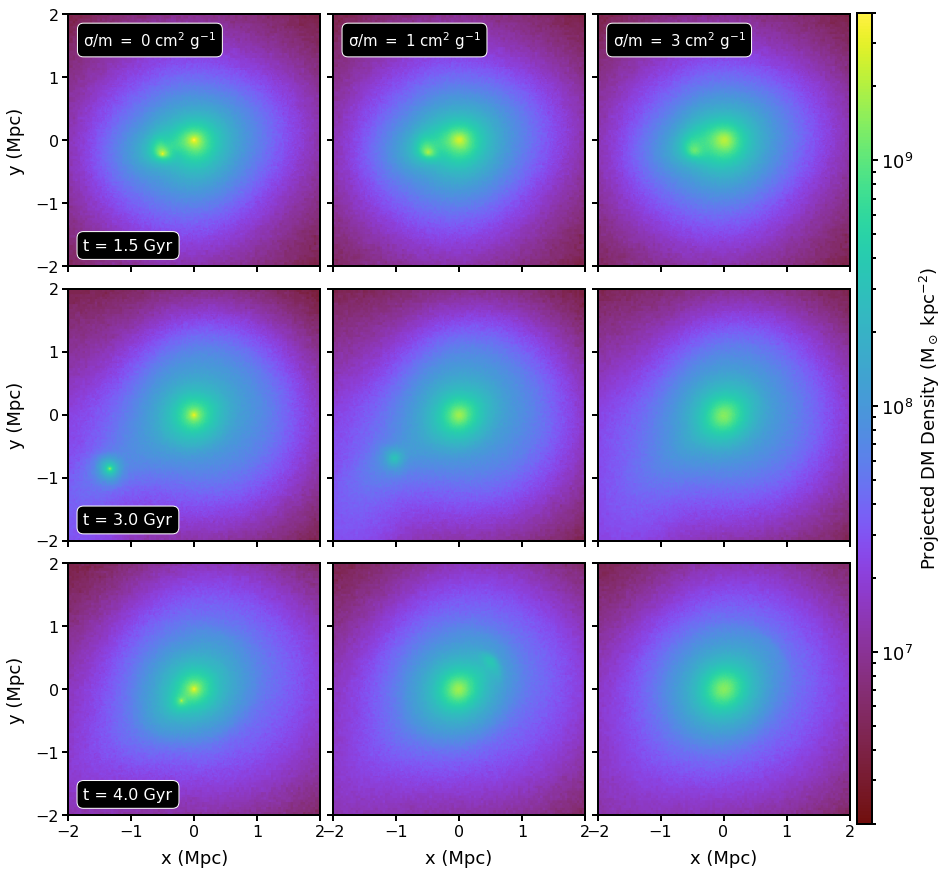}
\caption{Projected DM density at the epochs $t$ = 1.5, 3.0, and 4.0~Gyr for three different values of $\sigma/m$. Each panel is 4~Mpc on a side.}
\label{fig:dm_proj}
\end{figure*}

As mentioned above, the subcluster makes a second core passage at $t \approx$~3.6-4.1~Gyr. In the simulations with $\sigma/m \leq 1$~cm$^2$~g$^{-1}$, the subcluster survives as a more or less coherent structure. In the simulations with $\sigma/m$ = 3 and 10~cm$^2$~g$^{-1}$, collisions are so frequent that the subcluster loses its coherent structure shortly after the core passage and becomes a stream of particles within the main cluster's DM. Even in the $\sigma/m = 1$~cm$^2$~g$^{-1}$ case, the subcluster is undergoing a complete disruption following the second core passage. The evolution of the subcluster for simulations with different values of the DM cross section is illustrated in Figure \ref{fig:dm_proj}, which shows the projected DM density at three epochs following the core passage.

If there is no second core passage, then there will not be a second perturbation of the subcluster on the main cluster core. Since in our simulations the subcluster is gas-free, this second passage has a minor effect on the subsequent evolution of the cold fronts. More intriguingly, it is of note that in most observational accounts of sloshing cold fronts, identifying the subcluster which produced the original perturbation (typically via finding a second, smaller X-ray peak or a clump of galaxies) is often difficult. Examples include Virgo \citep{rod11} and A2204 \citep{chen2017}, though A1644 \citep{johnson2010} is a notable exception which has an obvious subcluster candidate, though the stage of the sloshing motions appears very early in this case. Our results show that for a non-negligible but observationally permitted values of the DM cross section ($\sigma/m \simlt 1$~cm$^2$~g$^{-1}$) small subclusters (and their associated gas and galaxies) may become somewhat subsumed into the main cluster after the second core passage, providing a partial explanation for the difficulty of identifying them in observations. However, a systematic study of the optical components of clusters with sloshing cold fronts is required before any conclusions can be definitively made. 
    
\subsubsection{Separation Between X-ray and SZ Peaks Due to DM Self-Interactions}\label{sec:xray_sz_sep}

Since sloshing motions are subsonic, they have a minor effect on the pressure profile of the gas in cool-core clusters, and the pressure peak remains very close to the potential miminum. Since the thermal Sunyaev-Zeldovich effect (hereafter tSZ) is a measure of the integrated pressure along the line of sight, the tSZ signal should be relatively unaffected by sloshing. This is seen clearly in the recent work on RXJ1347 by \citet{ueda2018}. However, the X-ray peak, which traces the densest and lowest-entropy gas in these systems, will get displaced from the cluster potential minimum. As we have already seen, this displacement reaches larger radii in simulations with non-zero DM cross section. 

\begin{figure*}
\centering
\includegraphics[width=0.95\textwidth]{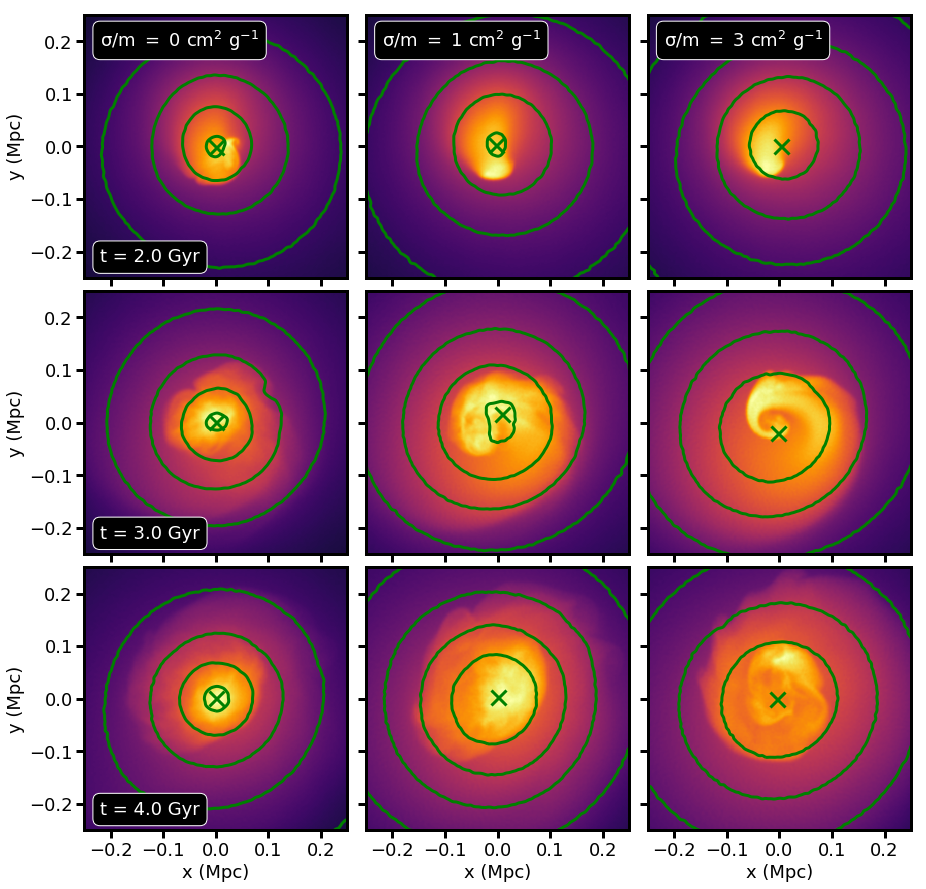}
\caption{Projected X-ray emissivity in the 0.5-7~keV band (in arbitrary units) with $y_{\rm tSZ}$ contours overlaid for three different values of $\sigma/m$ at the epochs $t$ = 2, 3, and 4~Gyr. The green ``$\times$'' symbol marks the position of the tSZ peak. Each panel is 0.5~Mpc on a side. The color scales for the X-ray emission are different in each panel and are chosen to show clearly the morphology.}
\label{fig:peak_sep}
\end{figure*}

In Figure \ref{fig:peak_sep}, we show maps of projected X-ray emissivity (in the 0.5-7~keV band) with contours of the Compton tSZ parameter $y_{\rm tSZ}$ overlaid for three epochs and three values of the DM cross section, which is defined by
\begin{equation}
y_{\rm tSZ} = \int\frac{k_BT}{m_ec^2}n_e\sigma_Td\ell
\end{equation}
The green ``$\times$'' in each panel marks the position of the tSZ peak. Shortly after the beginning of the sloshing process ($t = 2$~Gyr), all of the simulations exhibit a separation between the X-ray and tSZ peak of $\sim$20-40~kpc. However, the non-zero DM cross section simulations allow for greater and more long-lived separations between the tSZ and X-ray peaks. Without self-interactions, the two peaks already again coincide by $t = 3$~Gyr, but in the most extreme case shown of $\sigma/m = 3$~cm$^2$~g$^{-1}$, a separation between the two peaks of $\sim$80~kpc persists even to $t$ = 4~Gyr. This implies that for a given cluster a separation between these two peaks may provide an independent way to constrain the value of $\sigma/m$. This, however, would require knowledge of the stage of the sloshing motions, and would require a numerical simulation dedicated to matching the conditions of a particular cluster. Alternatively, this question could be addressed by running a large number of simulations over a wide parameter space in self-interaction cross section, mass ratio, and impact parameter, which would place tighter constraints on such X-ray/SZ separations due to SIDM.
                    
\section{Summary}\label{sec:summary}

We have performed a suite of simulations of core gas sloshing in a galaxy cluster core, building on previous work by adding the effect of DM self-interactions. The key ingredient in forming sloshing cold fronts in cluster cores via interactions with smaller clusters is the radically different collisionalities of the DM and baryonic components. Thus, the effect of a small DM cross section could potentially produce observable consequences on the formation and evolution of these features. Our main results are:

\begin{figure*}
\centering
\includegraphics[width=0.95\textwidth]{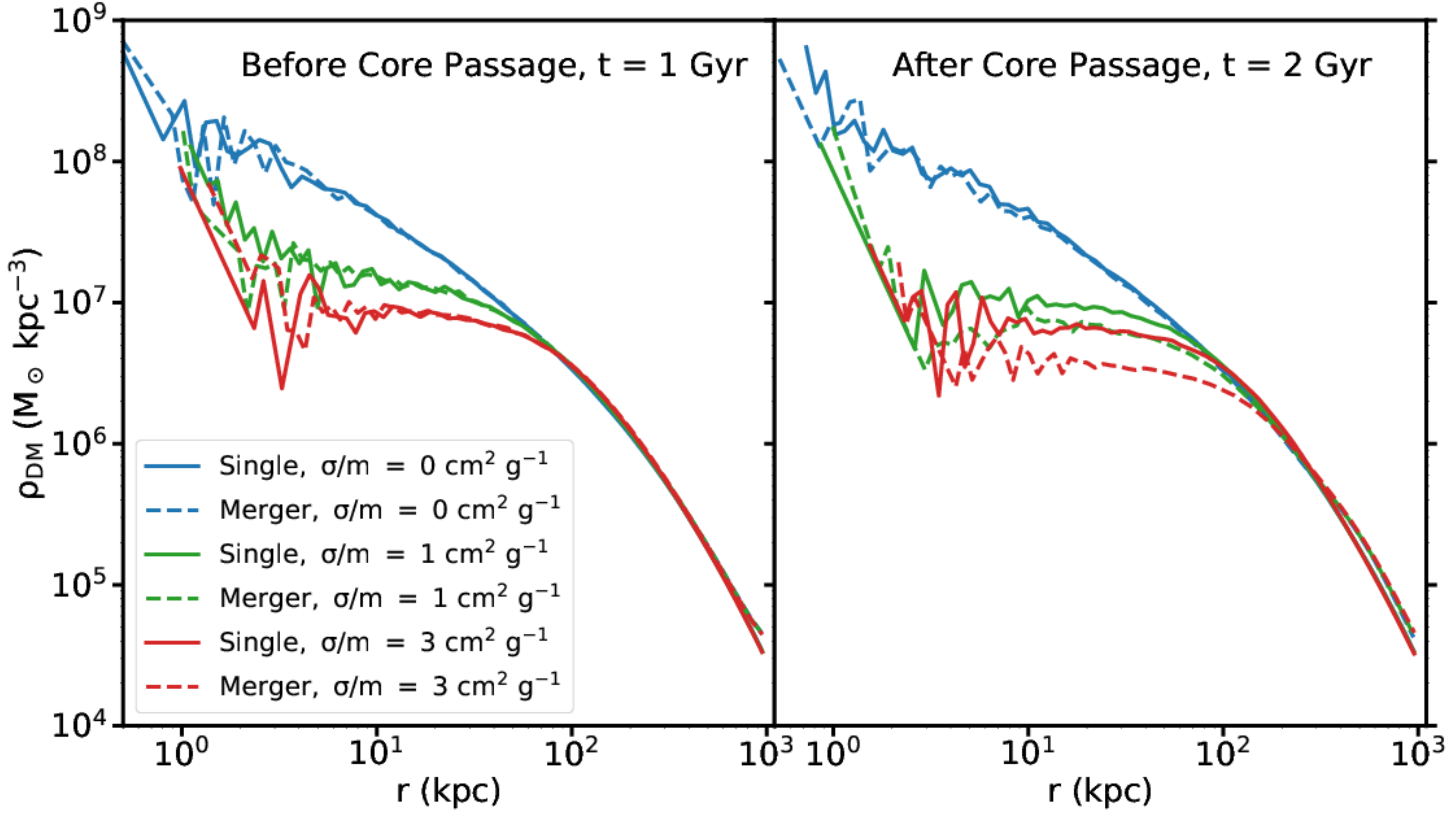}
\caption{Spherically averaged profiles of the DM density centered on the main cluster potential minimum before and after core passage for the single-cluster and merger runs for three different values of the DM cross section.}
\label{fig:before_after_density}
\end{figure*}
    
\begin{itemize}

\item In agreement with previous works, isolated cool-core clusters with initially cuspy DM density profiles gradually evolve flatter DM cores via DM self-interactions. The resulting gradual change in the gravitational potential causes a slow, adiabatic expansion and cooling of the gas in the center of the cluster. These changes are modest, so the essential thermodynamic structure of the cluster remains intact.

\item Sloshing cold fronts form in the same manner when the DM cross-section is non-zero as in the collisionless case. Due to the adiabatic cooling of the gas from the softening of the core, the sloshing gas is colder in simulations with larger DM cross-section. The cold fronts in simulations with signficant self-interactions are also less susceptible to the effects of KHI and turbulent mixing, at least in the earlier stages. 

\item In the earliest stages, the flattening of the potential caused by self-interactions enables the lowest-entropy gas to reach larger radii since there is a shallower potential to climb. On the other hand, because of the frequent and high-speed collisions the DM within subcluster experiences upon infall, it experiences further flattening of its potential and mass loss, and thus its influence on the core gas is weakened. The result is that the speed of the sloshing motions is therefore slower in the presence of self-interactions, which explains the slower growth of KHI in these simulations. The slower growth of turbulence and instabilities results in less turbulent mixing within the core region, and hence the lowest-entropy gas is longer-lasting. 

\item Large impact parameter-encounters with small subclusters do not produce a significant additional flattening of the larger cluster's core DM profile beyond what already has occurred due to its own self-interactions. However, interactions with the larger cluster's DM particles strip the subcluster of its DM and mix it in with the main cluster's DM on shorter timescales than would otherwise occur due to dynamical friction alone. 

\item The flattening of the DM core by self-interactions can result in significant separations between X-ray and SZ peaks which can persist for a number of Gyr. 

\end{itemize}

These conclusions lead to the obvious question--may they be used to constrain the value of $\sigma/m$ on cluster scales? The slower growth of KHI in the presence of DM self-interactions compared to the same merger scenario as collisionless DM is an intriguing result, especially considering most sloshing cold fronts appear to be relatively smooth. The reduced presence of KHI in observations with respect to their ubiquity in hydrodynamic simulations of cold fronts has typically been explained by magnetic fields \citep{zuh11} or viscosity \citep{rod13}. Magnetic fields are observed in clusters, and the viscosity of the ICM is still unknown, so there are far too many uncertainties to make any definite conclusions in this regard. Since the flattening of the inner density profile allows for the cold gas to climb to higher radii, a large sample of clusters could be examined for trends in the central density slope versus the radial extent of the sloshing gas. As mentioned above, SIDM may provide a partial explanation for the difficulty in easily locating subclusters which initiate the sloshing process, but this would require a more extensive study of the galaxy and lensing maps around many clusters with sloshing cold fronts. Perhaps the most stringent constraints on the SIDM cross-section could be placed by investigating separations between X-ray and SZ peaks in a large sample of sloshing cluster cores using high-angular-resolution SZ experiments in conjunction with X-ray observations.

Our idealized setup comes with a number of limitations that must be noted. Our initial conditions, which begin with two single NFW-like cuspy dark matter halos, are inherently out of equilibrium in the presence of self-interactions. This was evidenced by their transition to cored dark matter profiles in the run-up to the first core passage of the merger, and more or less remained in that state for the duration of the evolution of the cold fronts. This transition also resulted in adiabatic cooling and expansion of the gas, which had an effect on the temperature of the cold fronts, which was different in the different simulations. In a more realistic cosmological context, clusters are undergoing continuous accretion and merging with other clusters and groups, and the clusters themselves contain numerous smaller substructures down to galaxy scales. In this setting, the properties of the DM cores of clusters are likely to be more dynamic, becoming more core-like by self-interactions and more cusp-like via gravothermal collapse. Also, we have not included additional gas physics such as cooling, feedback, and star formation in these simulations. In these circumstances, gas would cool, condense, and form stars in the cluster center, which would have the effect of deepening the potential well and rendering the mass profile more cusp-like. A fully self-consistent picture would include all of these effects together, as in \citet{rob18a,rob18b}. Studying gas motions in the core of a cluster with DM self-interactions in such a cosmological context is left for future work.

It should also be noted that on longer timescales than those considered in this work SIDM halos eventually undergo a gravothermal catastrophe and recollapse, which would produce a cusp-like DM density profile in the central region \citep{kw2000}. This process will be sped up for inelastic collisions \citep{ess18}. Under such conditions, the potential would evolve during the cluster merger on a faster timescale and the effects on the ICM are likely to be more complex. 

Future papers will investigate the effects of DM self-interactions on the X-ray emitting plasma in major as well as minor mergers, examining both the thermodynamic and kinematic properties of the latter during the merger.

\acknowledgments
This work required the use and integration of a number of Python software packages for science, including AstroPy \citep{ast13}\footnote{\url{http://www.astropy.org}}, Matplotlib \citep{hun07}\footnote{\url{http://matplotlib.org}}, NumPy \footnote{\url{http://www.numpy.org}}, SciPy \footnote{\url{http://www.scipy.org/scipylib/}}, and yt \citep{tur11}\footnote{\url{http://yt-project.org}}. We are thankful to the developers of these packages. We also thank Volker Springel for the use of the \code{AREPO} code. JAZ thanks David Barnes, Rahul Kannan, Federico Marinacci, and Hui Li for useful discussions. He also thanks Grant Tremblay for assistance with a last-minute fix to a figure. JAZ acknowledges support through Chandra Award Number G04-15088X issued by the Chandra X-ray Center, which is operated by the Smithsonian Astrophysical Observatory for and on behalf of NASA under contract NAS8-03060. JZ acknowledges support by a Grant of Excellence from the Icelandic Research Fund (grant number 173929$-$051). The numerical simulations were performed using the computational resources of the Advanced Supercomputing Division at NASA/Ames Research Center.

\appendix

\section{The Effect of Adding a Stellar Component to the Subcluster}\label{sec:appendix}

As in previous works \citep[e.g.][]{AM06,zuh10,zuh11,rod13}, we used a gasless subcluster to initiate the sloshing process, which provides the cleanest setup to study the formation of cold fronts and their associated motions. To be consistent with our previous simulation investigations, the subcluster also lacks any stellar component. However, the central regions of relaxed galaxy clusters are dominated by BCGs \citep[e.g.][]{newman2013a,newman2013b}, and the stellar component of the mass will behave collisionlessly. This concentration of mass may deepen the subcluster potential enough to reduce the stripping of DM mass from the subcluster \citep[see, e.g.][]{Armitage2018}. 

To address this possibility, we have performed two simulations with identical initial conditions to the others, but with a BCG added to the subcluster. Using Equation \ref{eqn:sersic}, we add a galaxy with a mass of $M_{\rm{*}} = 1.2 \times 10^{12}~M_\odot$, which is appropriate for our subcluster mass of $2.5 \times 10^{14}~M_\odot$ \citep{kravtsov18}. Our two new simulations have DM cross sections of $\sigma/m = 0$~cm$^2$~g$^{-1}$ and $\sigma/m = 3$~cm$^2$~g$^{-1}$, respectively.

\begin{figure*}
\begin{center}
\includegraphics[width=0.9\textwidth]{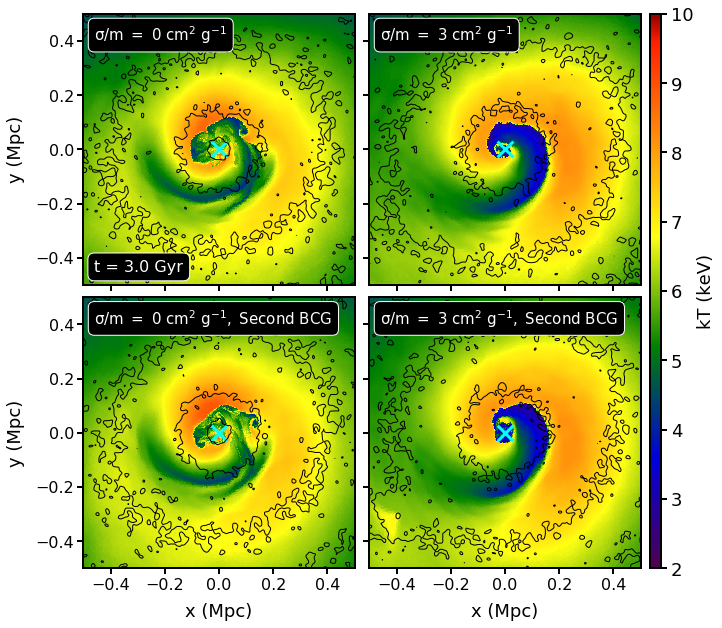}
\caption{Slices through the gas temperature in keV at the epoch $t$ = 3.0~Gyr for the merger simulations in which add a BCG to the subcluster. Contours are of dark matter density and are spaced logarithmically. The cyan ``$\times$'' marks the position of the center of mass of the BCG of the main cluster. Each panel is 1~Mpc on a side.}
\label{fig:mergers_bcg_3Gyr}
\end{center}
\end{figure*}

\begin{figure*}
\centering
\includegraphics[width=0.98\textwidth]{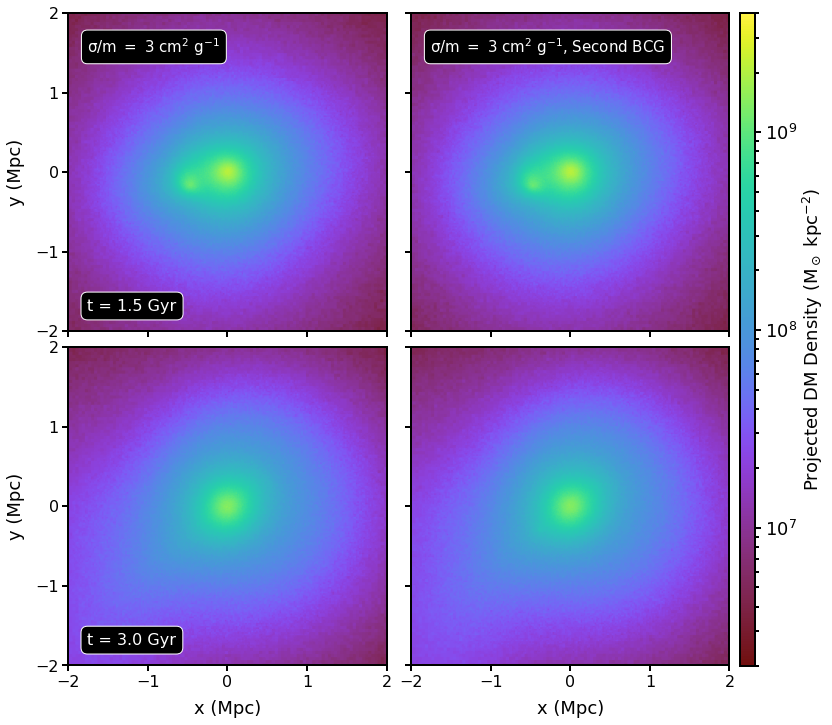}
\caption{Projected DM density at the epochs $t$ = 1.5, 3.0, and 4.0~Gyr for two simulations with $\sigma/m = 3$~cm$^2$~g$^{-1}$, where the subcluster is with (right panel) or without (left panel) a BCG. Each panel is 4~Mpc on a side.}
\label{fig:dm_proj_bcg}
\end{figure*}

Figure \ref{fig:mergers_bcg_3Gyr} shows the appearance of the cold fronts at $t = 3$~Gyr for both simulations where the subcluster  has a BCG. The behavior is the same as in our previous simulations--cold fronts are more disturbed by instabilities and turbulence when there are no self-interactions. Figure \ref{fig:dm_proj_bcg} shows the projected DM density at the two epochs of $t$ = 1.5 and 3.0~Gyr for both of these simulations. The presence of the subcluster BCG in the second simulation does not change the evolution of the subcluster's DM distribution appreciably. Thus, both of these figures show that presence of the BCG in the subcluster does not have a significant effect on our conclusions. 

\end{document}